\documentclass[10pt]{iopart}

\usepackage[margin=1.2in]{geometry}

\usepackage{iopams}  
\usepackage{cite}

\usepackage{cancel}
\usepackage{graphicx}

\usepackage{comment}

\usepackage[dvipsnames]{xcolor}
  \expandafter\let\csname equation*\endcsname\relax
  \expandafter\let\csname endequation*\endcsname\relax
\usepackage{amsmath}

\usepackage[T1]{fontenc}

\begin{document}
\title{Ion Heat and Parallel Momentum Transport by Stochastic Magnetic Fields and Turbulence}
\author{Chang-Chun Chen$^1$, P. H. Diamond$^1$, S. M. Tobias$^2$}
\address{$^1$Department of Physics, University of California San Diego, La Jolla, CA, USA}
 
\address{$^2$Department of Applied Mathematics, University of Leeds, Leeds LS2 9JT, UK}

\ead{chc422@ucsd.edu }

\begin{indented}
\item[]\today
\end{indented}


\begin{abstract}

The theory of turbulent transport of parallel momentum and ion heat by the interaction of stochastic magnetic fields and turbulence is presented.
Attention is focused on determining the kinetic stress and the compressive energy flux.
A critical parameter is identified as the ratio of the turbulent scattering rate to the rate of parallel acoustic dispersion.
For the parameter large, the kinetic stress takes the form of a viscous stress.
For the parameter small, the quasilinear residual stress is recovered.
In practice, the viscous stress is the relevant form, and the quasilinear limit is not observable. 
This is the principal prediction of this paper.
A simple physical picture is developed and shown to recover the results of the detailed analysis. 

\end{abstract}
\vspace{2pc}
\noindent{\it Keywords}: Stochastic Fields, Nonlinear Transport, Magnetohydrodynamics, Fusion Plasma
\\
%
\submitto{\PPCF}
%
%


\ioptwocol 

\section{Introduction and Basic Physics}

Heat transport, momentum transport, and the formation of shear flows in a stochastic field has long been recognized as a fascinating though complex problem in fusion devices.
It is one of the classic `paradigm problems' of magnetic fusion physics and has stimulated the writing of many well-known papers, most notably Rosenbluth et. al 1966 \cite{Rosenbluth_1966} and Rechester \& Rosenbluth1978 \cite{Rosenbluth1978}.
In any relevant application, turbulence will co-exist with the stochastic field. 
This is especially true for L-mode plasmas with resonant magnetic perturbations RMP (before the L-H transition), where the predominantly electrostatic turbulence is strongest just before transition.
Hence, studies on stochastic-field-induced effects in presence of strong turbulence is of importance in fusion plasma.

The bulk of the previous works focus on \textit{electron thermal transport} in a stochastic magnetic field \cite{joseph2008calculation, schmitz2009resonant, ida2015flow,Ohtani_2021}---this is on account of the tiny electron inertia, which is thought to allow long-distance electron streaming along wandering field lines.
Then, the more recent awareness of the need to achieve both good confinement and good power handling (and boundary control) has driven a resurgence of interest in the stochastic-field-induced transport problem. 
Topics of interest include, but are not limited to:
\begin{itemize}
	\item L-H transition dynamics in a stochastic magnetic field, as produced by (RMPs) \cite{evans2003}.
	It is now well known that the application of  RMPs raises the transition threshold\cite{Leonard_1991,Gohil_2011,Kaye_2011,Ryter_2013,Mordijck_2015,Scannell_2015, Schmitz_2019,Kriete2020} while it `stochasticizes' the edge layer.
	\item Intrinsic rotation in a stochastic magnetic field, as for the H-mode pedestal torque with RMPs \cite{Rice_1999,Rice_2004,Diamond_2013,Wang2013,diamond2013overview}.
	\item Internal transport barrier transitions triggered by magnetic islands\cite{CHIRIKOV1979263,Ida2001,Wolf_2002, Joffrin_2003,Ida_2010,Ida_2018}. 
\end{itemize}
Note that most or all of these phenomena are rooted in \textit{ion} transport and flow physics---topics rarely associated with the interaction between stochastic magnetic fields and turbulence.
This stochastic-field-induced effect was first analytically investigated by Chen et al. \cite{Chen_2020} which presented a theory of poloidal momentum transport induced by stochastic magnetic fields---the critical rate of stochastic-field-induced scattering $k_\perp v_A D_M$ required to dephase the turbulent poloidal Reynolds stress $\langle \widetilde{u}_r \widetilde{u}_\theta \rangle$ was calculated; here $k_\perp$ is wavenumber perpendicular to the mean field, $v_A$ is the Alfv\'enic speed, $D_M = \sum\limits_k |\widetilde{b}_k|^2  \pi \delta(k_\parallel) $ is the magnetic line diffusivity \cite{Rosenbluth_1966}, and $\widetilde{u}_r$, $\widetilde{u}_\theta$ are the perturbed radial and poloidal $E\times B$ flow velocity, respectively.
In this paper, we define $\widetilde{b}$ to be a root-mean-square (rms) of normalized fluctuating fields, i.e. $\widetilde{b} \equiv \sqrt{  \langle \widetilde{B} ^2 \rangle /B_0^2 }$, where $B_0$ is the mean toroidal magnetic field and the bracket average is an ensemble average over symmetry directions, i.e. $\langle \rangle \equiv \frac{1}{2\pi r} \int r d\theta \frac{1}{L_\parallel} \int dL_\parallel$.
Note that the Reynolds stress and force are related to the vorticity flux by the Taylor identity\cite{taylor1915}, and that the vorticity flux enters $\nabla \cdot J =0$.
The Alfv\'en speed $v_A$ then emerges as the speed characteristic of the decorrelation process here. 
The competition of stochastic field scattering and ambient turbulent decorrelation determines the field fluctuation intensity $\widetilde{b}^2$ which can suppress the transition, or equivalently, the increment in power needed to transition in the presence of $\widetilde{b}^2$.
However, a moment's consideration of the ion radial force balance equation
\[
	\langle E_r \rangle = \frac{\nabla \langle p_i\rangle}{ne} - \langle \bold{u} \rangle \times \langle \bold{B} \rangle,
\]
reminds us that in addition to mean poloidal flow $\langle u_\theta \rangle$, the evolution of mean parallel flow $\langle u_\parallel \rangle$ and ion pressure $\langle p_i \rangle$ should also be revisited in the context of \textit{co-existing} backgrounds of turbulence and stochastic magnetic fields. 
To this end, this paper addresses aspects of ion energy and parallel momentum transport induced by the \textit{interaction} of stochastic fields and turbulence. 


Motivated by studies of rotation damping due to ergodic magnetic limiter operation on the TEXT \cite{yang1991space}, Finn et al. \cite{finn1992} (hereafter referred to as FGC) addressed the `stochastic field only' limit of the problem.
The FGC analysis begins from the mean field evolution equation of the parallel flow and pressure ---
\begin{equation}
	\frac{\partial}{\partial t} \langle u_\parallel \rangle 
	+\frac{\partial}{\partial r}  \langle \widetilde{u}_r \widetilde{u}_\parallel \rangle 
	= -\frac{1}{\rho} \frac{\partial}{\partial r} \langle \widetilde{b}_r, \widetilde{p}\rangle 
\end{equation}
\begin{equation}
	\frac{\partial}{\partial t} \langle p \rangle  
	+ \frac{\partial}{\partial r}  \langle \widetilde{u}_r \widetilde{p} \rangle
	=
	-\rho c_s^2 \frac{\partial}{\partial r} \langle \widetilde{b}_r \widetilde{u}_\parallel \rangle ,
\end{equation}
where $c_s \equiv \sqrt{\gamma p/\rho}$ is the sound speed, $\gamma$ is the adiabatic index, and $\rho$ is the mass density.
The familiar advective fluxes of the parallel flow and pressure are ignored. 
Our goal is then  to calculate the kinetic stress ($K \equiv  \langle \widetilde{b}_r \widetilde{p} \rangle / \rho$) and the compressive energy flux ($H \equiv \rho c_s^2 \langle \widetilde{b}_r \widetilde{u}_\parallel \rangle$). 
Note that the divergence of the kinetic stress $\partial_r K$ drives mean parallel flow $ \langle u_\parallel \rangle$ via the pressure gradient along tilted magnetic field lines, while the divergence of the compressive energy flux  $\partial_r H$ couples field line tilting to compressive heat flow so as to drive energy transport.
We note in passing that the kinetic stress has been linked directly to plasma rotation by studies on the Madison Symmetric Torus reverse field pinch \cite{Ding2013,Sarff_2013}.
By a combination of probes and polarimetry, Ding et al.\cite{Ding2013} demonstrated a clear correlation between the divergence of the measured kinetic stress and the mean $\langle u_\parallel \rangle$ profile (see Figure 2. of Ding et al.\cite{Ding2013}).
This result establishes that stochastic magnetic fields can impact flow dynamics. 
It is also a compelling insight into the connection among fluctuation measurements, parallel flow dynamics, and momentum transport.
Hence, Ding's study constitutes a rare link between the microscopic and macroscopic facets of the momentum transport problem.

Given the clear resemblance of this problem to aspects of gas dynamics, a natural approach is to cast the analysis in terms of the familiar Riemann variables $u_\parallel \pm p$ \cite{LL_fluid1959}.
A quasilinear analysis then gives an estimate of the relaxation rate for excitation on a perpendicular scale length $l_{\perp}$ as $c_s D_M/l_\perp^2$. 
This rate may be thought of as characteristic of acoustic pulse decorrelation due to propagation along stochastic field lines. 
However, it should be said that the dynamics here are fundamentally \textit{non-diffusive}. 
In particular, the kinetic stress (i.e. $K = -c_s D_M \partial_r \langle p\rangle $) actually is \textit{residual stress} driven by $\nabla \langle p \rangle$\cite{Gurcan2010}.
Likewise, the compressive energy flux (i.e. $H = -c_s D_M \partial_r \langle  u_\parallel \rangle $) is a non-diffusive contribution to the energy flux driven by $\nabla \langle u_\parallel \rangle $---this may be thought of like a pinch.
These relations were not presented in  FGC.
Also, we observe that since basic physics is fundamentally one of a \textit{static} stochastic field, all key results may be obtained by working directly with $\bold{B}\cdot \nabla p =0$ and $\bold{B}\cdot \nabla u_\parallel =0$.


The analysis discussed so far was quasilinear. 
FGC obtained expressions for kinetic stress  ($K = \langle \widetilde{b}_r \widetilde{p} \rangle /\rho$) and compressive energy flux ($H = \rho c_s^2 \langle \widetilde{b}_r \widetilde{u}_\parallel \rangle$) by computing responses $\widetilde{p} \propto (\delta p/\delta b) \widetilde{b} $,  $\widetilde{u}_\parallel \propto (\delta u_\parallel /\delta b) \widetilde{b} $ and closing the expressions for the kinetic stress and compressional heat flux, yielding both proportional to $ D_M = \sum\limits_k | \widetilde{b}_k|^2 l_{ac}$.
Here, $\delta p/\delta b$ and $\delta u_\parallel/\delta b$ are responses of pressure and parallel flow to the magnetic perturbation $\widetilde{b}$.
The issue is the assumption concerning the physics content of the responses $\delta p/\delta b$ and $\delta u_\parallel/\delta b$.
To address this turbulent limit, one must calculate the kinetic stress and compressional heat flux \textit{in the presence of electrostatic turbulence}---i.e. the responses $\delta p/\delta b$ and $\delta u_\parallel/\delta b$ must be computed in the presence of a scattering field of electrostatic fluctuations, which we represent as a spectrum of fluctuating $E \times B$ velocities $\langle \widetilde{\bold{u}} \; \widetilde{\bold{u}} \rangle_{k,\omega}$.
As we will show, this makes for a significant and \textit{qualitative} departure from the quasilinear analysis. 
Note that this analysis is, in some sense, `dual' to that of Chen et al.\cite{Chen_2021}.
There, the vorticity response $\delta U/\delta \widetilde{u}_r$ was calculated in the presence of a prescribed ensemble of $\langle \widetilde{b}^2 \rangle $ and used to calculate the Reynolds stress, where $U = \nabla_\perp^2 \widetilde{\phi}/B_0$ is the $E\times B$ vorticity and $\phi$ denotes the electrical potential.
Here, we compute the pressure and parallel flow responses $\delta p/\delta b$ and $\delta u_\parallel /\delta b$ in the presence of electrostatic turbulence and use them to calculate the kinetic stress component $\langle \widetilde{b} \widetilde{p} \rangle$.
Implicit in both is the assumption that the statistics of the magnetic perturbation field causing the stochasticity are independent of those of the electrostatic perturbation field of the turbulence, i.e. we assume $\langle \widetilde{b} \widetilde{\phi}  \rangle =0$.
This ansatz eliminates cross-terms from the calculations of fluxes. 
We discuss this assumption further, later in the paper. 
 
\begin{figure}[h!]
\centering
  \includegraphics[scale=0.18]{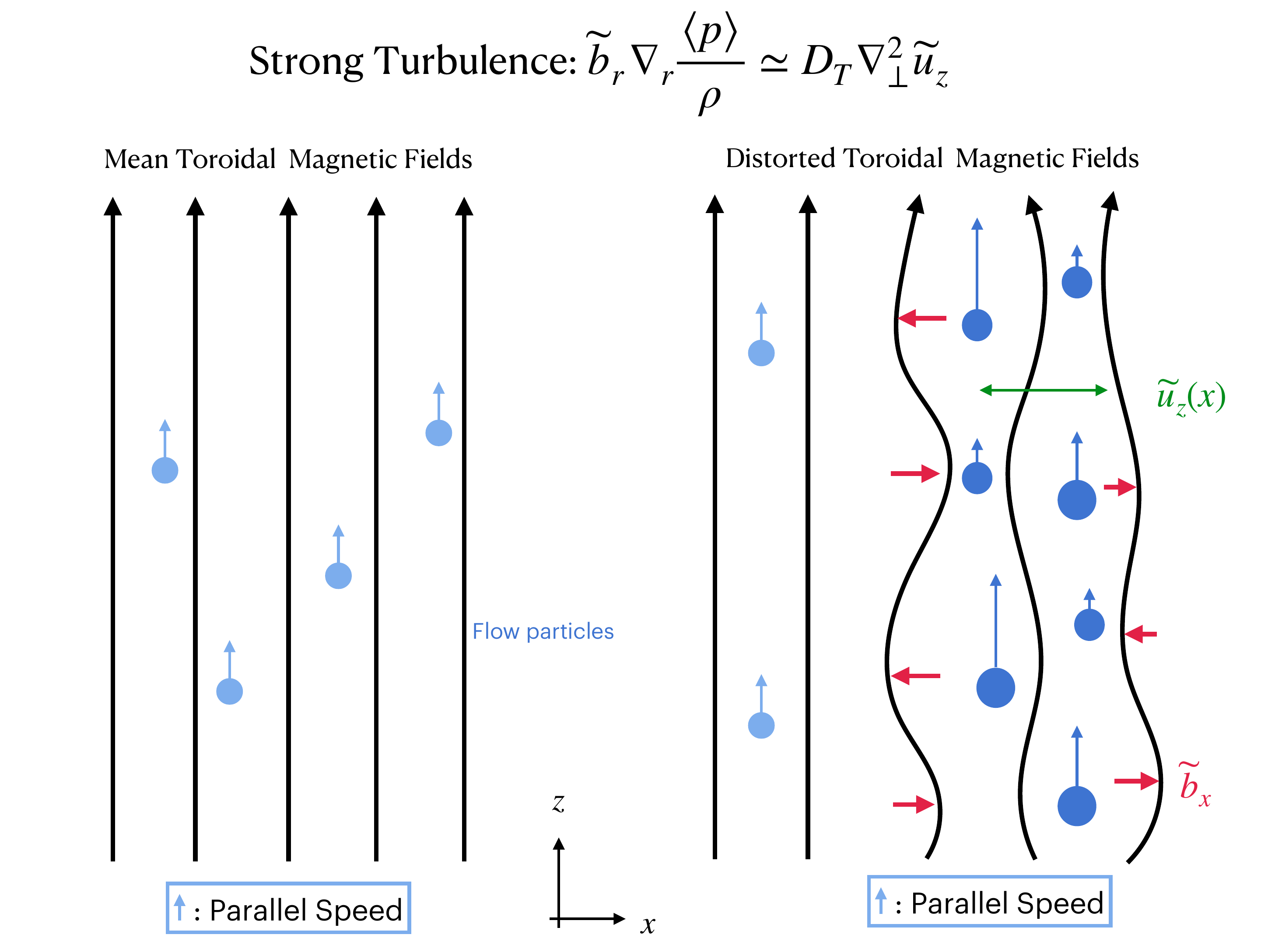}
  \caption{Left: Mean magnetic field in the parallel direction with fluid flow speed. Right: The magnetic field is perturbed by the stochastic field $\widetilde{b}_r$.
  In response to magnetic perturbation and pressure slug, turbulent viscosity $\nu_T$ will dissipate the parallel flow perturbation.
   In \Sref{sec: Strong turb}, we obtain that the change in mean pressure ($\partial_x \langle p \rangle/ \rho $)  is balanced by turbulent mixing of parallel flow, i.e. $\nabla_\perp^2 \widetilde{ u}_z $. 
   Blue arrows indicate the change of parallel speed.}
  \label{fig:Strong_turb_flow}
\end{figure}

\begin{figure}[h!]
\centering
  \includegraphics[scale=0.18]{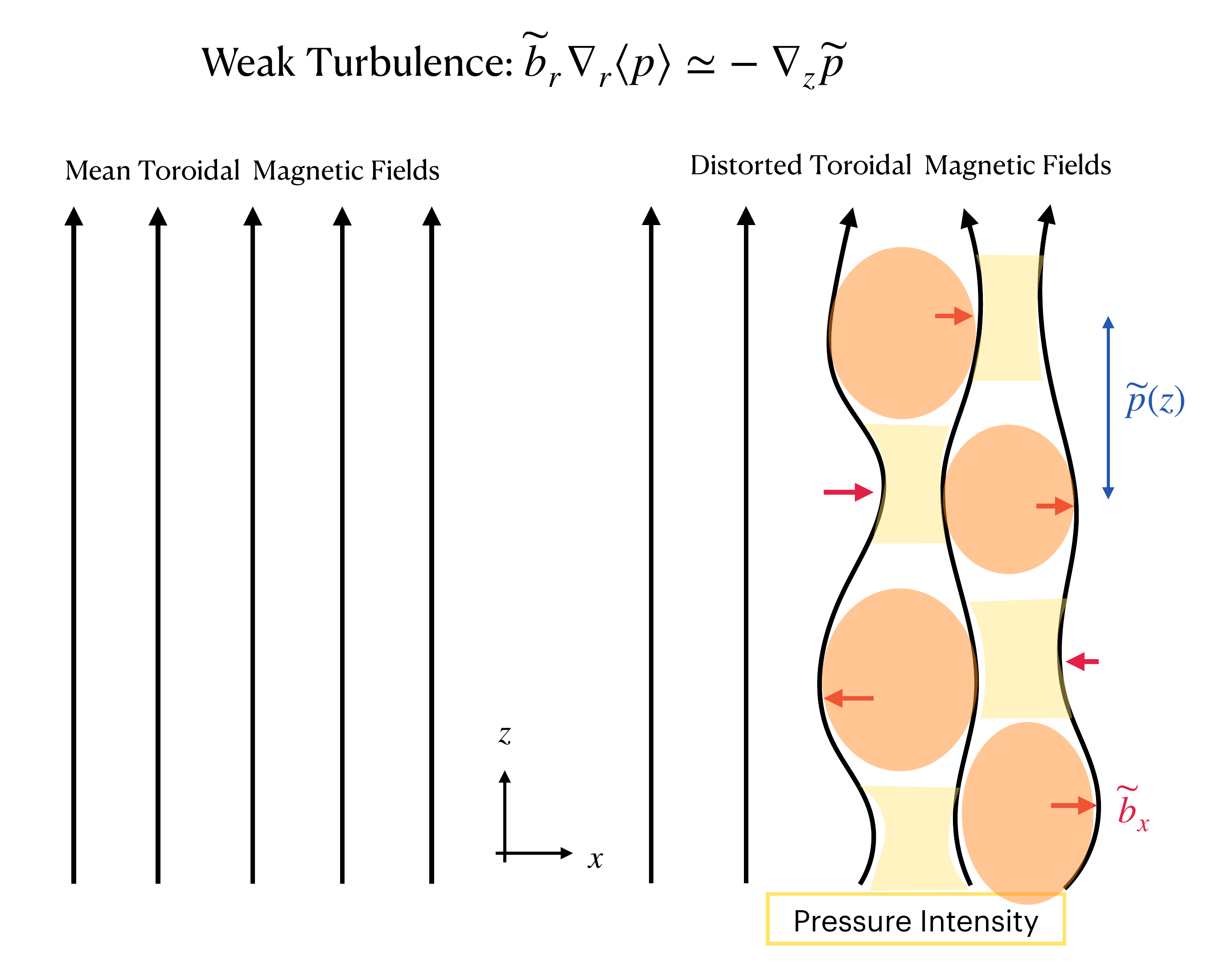}
  \caption{Left: Mean magnetic field in toroidal direction with constant pressure in $z$-direction.
  Right: The magnetic field is perturbed by the stochastic field $\widetilde{b}_r$.
   In response to magnetic perturbation and pressure slug, the pressure gradient will build up along the mean field.
   Regions with higher and lower pressure intensity are colored orange and yellow, respectively.
  }
  \label{fig:Weak_turb_pressure}
\end{figure}

A heuristic but enlightening model of the pressure response $\delta p/\delta b$ is presented here and serves to guide the reader through the subsequent detailed analysis. 
The parallel flow response $\delta u_\parallel/\delta b$ can be estimated in a similar way.
Hence, we discuss only $\delta p/\delta b$. 
Here, it is helpful for the reader to consult  \Fref{fig:Strong_turb_flow} and \ref{fig:Weak_turb_pressure}.
One can `pluck' a magnetic field line by $\widetilde{b}$. 
Since a mean radial pressure gradient $\partial_r \langle p\rangle$ is present, the magnetic perturbation will generate a localized slug of pressure excess-per-length $\widetilde{b}_r \partial_r \langle p\rangle$.
To balance this local pressure excess, there are two possibilities:
\begin{itemize}
	\item If the rate of turbulent (i.e. viscous) mixing of the parallel flow response is large (i.e. $\nu_T/l_\perp^2>$ other rates), then a turbulent viscosity $\nu_T$ will dissipate the parallel flow perturbation $\widetilde{u}_\parallel$, produced in response to the magnetic perturbation and pressure slug (see \Fref{fig:Strong_turb_flow}).
	In this case, $\nu_T \nabla_\perp^2 \widetilde{u}_\parallel \simeq  \widetilde{b}_r \partial_r \langle p\rangle/\rho$, where $\nu_T$ is the turbulent viscosity due to the electrostatic turbulence. 
	In this limit, perturbed pressure is replaced by a dynamic balance of the turbulent Reynolds force with the local pressure excess. 
	Here, $\nu_T \simeq D_T \simeq \int dt^\prime \langle \widetilde{u}_{\perp} (0) \widetilde{u}_{\perp} (t^\prime)  \rangle  \simeq \langle \widetilde{u}_\perp^2 \rangle \tau_{ac}$, where $\tau_{ac}$ is the autocorrelation time of the electrostatic fluctuation and $D_T$ is the turbulent fluid diffusivity. 
	\item If the rate of sound propagation along the perturbed field is large (i.e. $c_s/l_\parallel >$ other rates), then a pressure gradient will build up along the mean field, so as to cancel the initial imbalance due to the slug (see \Fref{fig:Weak_turb_pressure}).
	In this case, $\widetilde{p} c_s/l_\parallel \simeq -c_s \widetilde{b}_r \partial_r \langle p\rangle$, which leads to the quasilinear result for $\widetilde{p}$ and $K$. 
\end{itemize}

Here, the critical competition (highlighted in \Fref{fig:Weak_turb_pressure} and \Fref{fig:Strong_turb_flow}) is that between the parallel acoustic transit rate $c_s/l_\parallel \simeq c_s \Delta k_\parallel$ and the perpendicular diffusive mixing rate $\simeq \nu_T/l_\perp^2 \simeq k_\perp^2 D_T$.
Hereafter, we take $|k_\parallel| \simeq \Delta k_\parallel $, where $k_\parallel$ may change sign and $\Delta k_\parallel$ is always positive. 
In most relevant cases (i.e. as for drift wave turbulence), $k_\perp^2 D_T \simeq \omega > k_\parallel c_s$, so the dynamic balance regime is relevant.
Note that in this regime, the \textit{qualitative} form of the response to $\widetilde{b}$ differs from the quasilinear case. 
In particular, a hybrid viscous stress replaces the residual stress and involves turbulent decorrelation resulting from scattering by electrostatic fluctuation. 
In the weak turbulence regime, we recover perturbed pressure balance.  
The detailed analysis supports the conclusion derived from heuristics here.  

The remainder of this paper is organized as follows.
\Sref{sec: Model} presents the models and discusses the quasilinear theory. 
It also presents the \textit{explicit} calculation of particle flux and the parallel momentum transport in a steady electric field. 
\Sref{sec: kinetic stress} analyzes the physics of kinetic stress ($K$) and compressive energy flux ($H$), which play important roles in momentum and density evolution.
These are calculated in the presence of turbulence.
\Sref{sec: Discuss} discusses the applications of the theory, along with future work.


\section{Models and Transport by Static Stochastic Fields} \label{sec: Model}

\begin{figure}[h!]
\centering
  \includegraphics[scale=0.3]{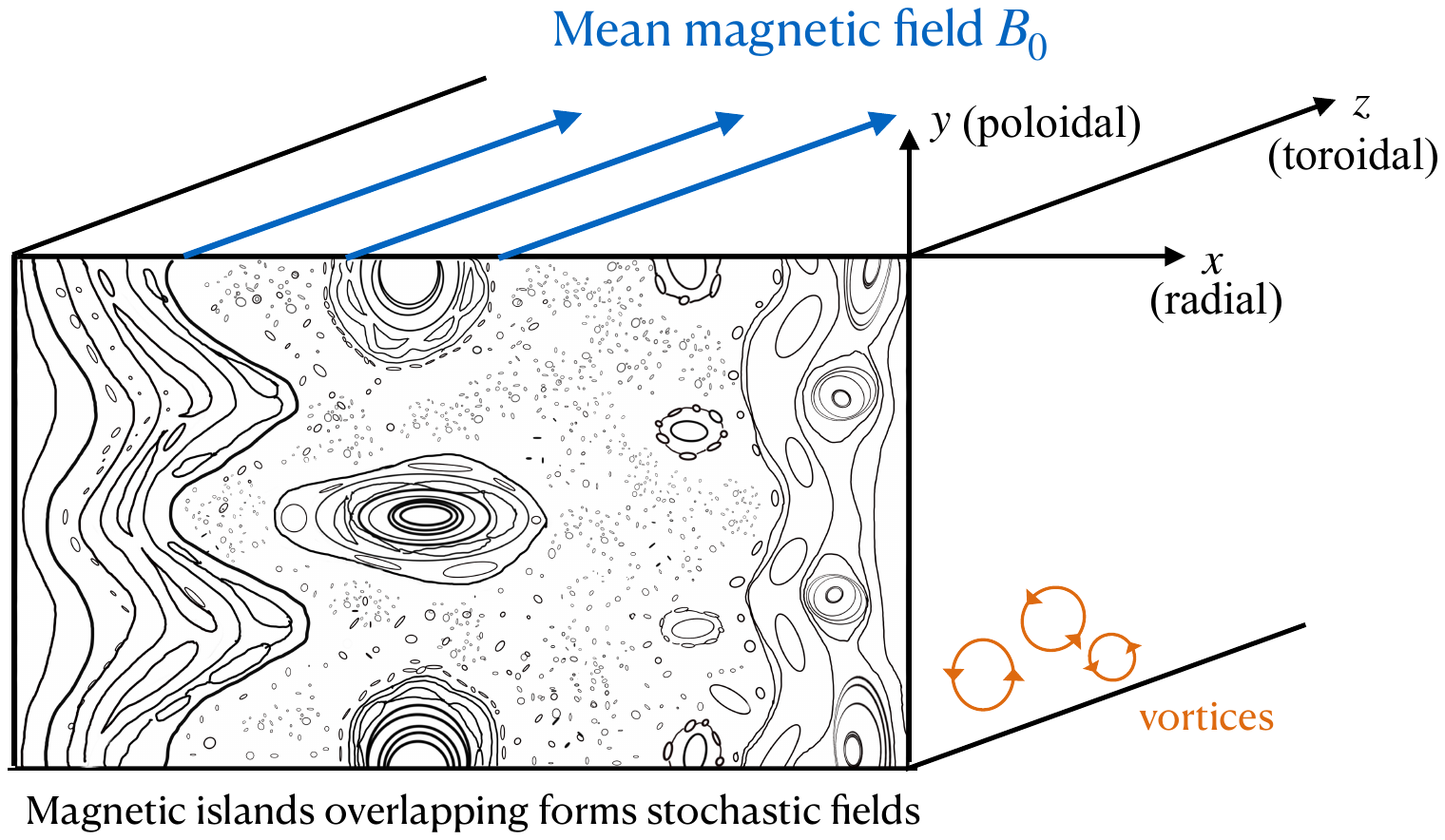}
  \caption{Magnetic fields at the edge of tokamak. RMPs-induced magnetic islands (black lines) lie in radial ($x$) and poloidal ($y$) plane.  
  Mean toroidal field is treading through stochastic fields perpendicular in $z$-direction (blue arrows).}
  \label{fig:3D}
\end{figure}

Here, we construct a model for the evolution of density and parallel flow in the presence of stochastic fields in Cartesian (slab) coordinates used in Chen et al. \cite{Chen_2021} --- $x$ is radial, $y$ is poloidal, and $z$ is toroidal direction, in which the mean toroidal field lies (\Fref{fig:3D}). 
In this 3D system, the stochasticity of magnetic fields, given by a response to an external excitation such as an RMP coil, results from the overlap of magnetic islands located at resonant $\underline{k}\cdot \underline{B} =0$ surfaces \cite{wesson2011tokamaks}. 
Once overlap occurs, the coherent character of the perturbations is lost, due to finite Kolmogorov-Sinai entropy (i.e. there exists a positive Lyapunov exponent for the field) \cite{kolmogorov1959entropy,sinai1959notion}. 
Hence, the total magnetic field can be decomposed into the mean toroidal (parallel) field on the $z$-axis plus the stochastic field lying in $x-y$ plane.
In this case, we take the magnetic Kubo number \cite{kubo1963} to be modest $Ku_{mag} \lesssim 1$ --- so mean field theory is valid. 
This is consistent with reported experimental values of magnetic perturbations $\widetilde{b}$ \cite{Yu_2000,Wolf_2005,evans2006edge,In_2015}.
We decompose the magnetic fields, magnetic potential, velocities, electric potential, pressure, and density
\begin{equation}
	\begin{cases}
		\text{magnetic fields} & \bold{B} = (\widetilde{B}_{x}  , \widetilde{B}_{y} , \; B_0 )
		\\
		\text{potential fields} & \bold{A} = (-\frac{1}{2} B_0 y, \; \frac{1}{2} B_0 x ,  \; \widetilde{A}_{(x,y)})
		\\
		\text{velocities} & \bold{u} = (\widetilde{u}_x,  \; \langle u_y \rangle +\widetilde{u}_y, \;  \langle u_z \rangle + \widetilde{u}_z)
		\\
		\text{electric potential} & \phi = \langle \phi \rangle  + \widetilde{\phi}
		\\
		\text{pressure} & p = \langle p \rangle  + \widetilde{p}
		\\
		\text{particle density} & n= \langle n \rangle  + \widetilde{n}.
	\end{cases}
\end{equation}
Here $\langle u_y \rangle$ is the mean poloidal flow, $ \langle u_z \rangle$ is the mean parallel flow. 
The tilde $\widetilde{\;\;}$ denotes the perturbations of the mean. 

\subsection{Non-diffusive Effect for Electron Particle Flux}
In this section, we discuss the transport of parallel momentum and particles due to stochastic fields. 
One aim here is to make contact with and clarify the FGC result \cite{finn1992} as a baseline for later studies of stochastic scattering along with turbulence.
Another is to elucidate the contribution to the physics of particle transport in a stochastic magnetic field --- i.e. to determine the physical significance of the result.
Following FGC, here, we assume an isothermal plasma, so the basic equations reduce to:
\begin{equation}
	n \frac{\partial}{\partial t} u_z
	= -c_s^2 \nabla_z n,
\end{equation}
\begin{equation}
	\frac{\partial}{\partial t} n 
	= -n \nabla_z u_z,
\end{equation}
where $\nabla_z =\nabla_z^{(0)} + \widetilde{b}\cdot \nabla_\perp$.
Then the mean fields $\langle u_z \rangle$ and $\langle n \rangle $ evolve according to 
\begin{equation}
	n_0 \frac{\partial}{\partial t} \langle u_z \rangle
	= -c_s^2 \frac{\partial}{\partial x}  \langle \widetilde{b}_x \widetilde{n}, \rangle
	\label{eq: sec.2.5 u_z evolution}
\end{equation}
\begin{equation}
	\frac{\partial}{\partial t} \langle n \rangle
	= -n_0 \frac{\partial}{\partial x}  \langle \widetilde{b}_x \widetilde{u}_z \rangle,
	\label{eq: sec.2.5 n evolution}
\end{equation}
where $n_0$ is a static, uniform background density.
Thus, determining the effect of a stochastic field on density evolution (i.e. particle transport) requires a calculation of the flux $ \langle \widetilde{b}_x \widetilde{u}_z \rangle$.
Likewise, for the effect on parallel flow, the kinetic stress $c_s^2 \langle \widetilde{b}_x \widetilde{n} \rangle$ is needed. 
The physical interpretation of how the density evolution discussed here is related to the particle flux is discussed at the end of this section.

To calculate $\langle \widetilde{b}_x \widetilde{u}_z \rangle$ and $\langle \widetilde{b}_x \widetilde{n} \rangle$, we proceed by quasilinear theory. 
Proceeding from \Eref{eq: sec.2.5 u_z evolution} and \eref{eq: sec.2.5 n evolution}, these equations can be written as 
\begin{equation}
	\frac{\partial}{\partial t}\frac{\widetilde{u}_z}{c_s} 
	= -c_s \nabla_z^{(0)} \frac{\widetilde{n}}{n_0} 
	- c_s\widetilde{b}_x \frac{\partial}{\partial x} \frac{\langle n\rangle }{n_0},
	\label{eq: finn 1}
\end{equation}
\begin{equation}
 	\frac{\partial}{\partial t}\frac{\widetilde{n}}{n_0} 
 	= -c_s \nabla_z^{(0)} \frac{\widetilde{u}_z}{c_s}
	 - c_s \widetilde{b}_x \frac{\partial}{\partial x} \frac{\langle 	u_z \rangle}{c_s} .
	\label{eq: finn 2}
\end{equation} 
We combine \Eref{eq: finn 1}$\pm$\eref{eq: finn 2} and obtain the Riemann-like variables $h_{\pm} \equiv \frac{\widetilde{u}_{z}}{c_s} \pm \frac{\widetilde{n}}{n_0}$, and the Riemann equation
\begin{equation}
	\frac{\partial}{\partial t} h_{\pm} \pm c_s \frac{\partial}{\partial z} h_{\pm} 
	= 
	- c_s\widetilde{b}_x \frac{\partial}{\partial x} \frac{\langle n\rangle}{n_0} \mp c_s \widetilde{b}_x \frac{\partial}{\partial x} \frac{\langle u_z \rangle }{c_s} .
	\label{eq: h_pm}
\end{equation}
Note that $h_\pm$ propagate at $c_s$ in opposite directions.
Now, the magnetic perturbations here are \textit{static}, so we can immediately take $\partial_t h_\pm =0$.
No acoustic wave dynamics enters, though the acoustic speed appears in the problem. 
From \Eref{eq: h_pm}, we can then immediately write 
\begin{equation}
	 h_{\pm, k} 
	 = \mp  \int dl 
	 [\widetilde{b}_x \frac{\partial}{\partial x} \frac{\langle n\rangle}{n_0}
	  \pm  \widetilde{b}_x \frac{\partial}{\partial x} \frac{\langle u_z \rangle }{c_s} 
	  ].
\end{equation}
Here, $l$ parameterizes the distance along a magnetic field line, and the solution of \Eref{eq: h_pm} is affected by integrating along static stochastic field lines. 
Now, $\widetilde{u}_{z}/c_s $ and $ \widetilde{n} / n_0$ can be recovered noting 
\begin{equation}
	\frac{\widetilde{u}_{z}}{c_s} 
	= (h_+ + h_-)/2
	= -\int dl \widetilde{b}_x \frac{\partial}{\partial x} \frac{\langle u_z \rangle }{c_s},
\end{equation}
\begin{equation}
	 \frac{\widetilde{n}}{n_0}
	= (h_+ - h_-)/2 
	= - \int dl \widetilde{b}_x \frac{\partial}{\partial x} \frac{\langle n\rangle}{n_0},
\end{equation}
It then follows that 
\begin{equation}
	\langle \widetilde{b}_x \widetilde{u}_z \rangle
	= -D_M \frac{\partial}{\partial x} \langle u_z \rangle,
	\label{eq: bu caused by D_M}
\end{equation}
\begin{equation}
	\langle \widetilde{b}_x \widetilde{n} \rangle
	= -D_M \frac{\partial}{\partial x} \langle n \rangle,
	\label{eq: bn caused by D_M}
\end{equation}
where 
\begin{equation}
	D_M 
	= \int dl \langle \widetilde{b}_x(0) \widetilde{b}_x(l) \rangle
	= \langle \widetilde{b}_x^2 \rangle l_{ac},
\end{equation}
and $\langle \widetilde{b}_x(0) \widetilde{b}_x(l) \rangle$ is the magnetic perturbation correlation function, $D_M$ is the usual stochastic field diffusivity, and $l_{ac}$ is magnetic perturbation auto-correlation length such that
\begin{align}
	l_{ac} = \frac{1}{\Delta k_\parallel}.
\end{align}
The mean field density equation, of this state, is then 
\begin{align}
	\frac{\partial}{\partial t} \frac{\langle n \rangle}{n_0}
	= \frac{\partial}{\partial x}  c_s D_M \frac{\partial}{\partial x} \frac{\langle u_z \rangle}{c_s}.
	\label{eq: mean evol. of n}
\end{align}
Note that other physical processes enter the full evolution of density, as discussed below.

Several aspects of these results merit some discussion.
First, \Eref{eq: bu caused by D_M} and \eref{eq: bn caused by D_M} are flux-gradient relations with characteristic transport coefficient $c_s D_M$. 
Thus the characteristic rate for perturbations on scale $l_\perp$ is $1/\tau(l_\perp) \simeq c_s D_M/l_\perp^2$, as noted by FGC.
However, the actual fluxes in \Eref{eq: bu caused by D_M} and \eref{eq: bn caused by D_M} are not diffusive, but rather off-diagonal, leading to cross-coupling of $\langle n\rangle$ and $\langle u_z\rangle$ evolution.
In particular, $\langle \widetilde{b}_x \widetilde{u}_z \rangle$ yields an off-diagonal convective flux, not particle diffusion.
Likewise, $\langle \widetilde{b}_x \widetilde{n} \rangle$ contributes a fundamentally non-diffusive \textit{residual stress}, but not a viscosity.
FGC overlooked these points since that their analysis never transformed back from Riemann variables (referred to as Els\"{a}sser variables by FGC) to physical variables.
We note also that the results of \Eref{eq: bu caused by D_M} and \eref{eq: bn caused by D_M} may be obtained directly from linearizing 
\begin{align}
	B \cdot  \nabla u_z =0
	\\
	B \cdot  \nabla n =0
\end{align}
and using $\widetilde{u}_z$ and $\widetilde{n}$ to derive the fluxes. 
The problem is fundamentally static, and no sound wave dynamics is involved.


\Eref{eq: mean evol. of n} describes the piece of density evolution in a stochastic magnetic field which is due to $c_s \widetilde{b} \cdot \nabla \langle u_z \rangle$.
A natural question is how is this related to the full particle flux, as it is conventionally understood. 
FGC refers to the density evolution in this problem as `sound wave transport', yet it is clear that no sound wave dynamics is involved. 
To clarify this question, we consider electron particle transport in a stochastic magnetic field. 
Here, we consider the stochastic field as co-existing with plasma current perturbations which generate it, so that Amp\`ere's law is satisfied. 
This does not preclude the possibility of external excitation of the stochasticity, as by an RMP. 
The drift kinetic equation for electrons is 
\begin{equation}
	\frac{\partial}{\partial t} f+ u_z  \cdot \nabla_z f - \frac{E_\perp \times \hat{z}}{B_0} \cdot \nabla f
	-\frac{|e|}{m} E_z \frac{\partial}{\partial u_z} f = C(f),
\end{equation}
so that the electron density evolution due to $\widetilde{b}$ effects is determined by 
\begin{equation}
	\frac{\partial n_e}{\partial t} = - n_{0,e} \nabla_z u_{z,e},
\end{equation}
where $f$ is a general distribution function and $u_{z,e}$ is the parallel electron flow. 
Then for mean electron density, noting that $\nabla_z =\nabla_z^{(0)} + \widetilde{b}\cdot \nabla_\perp$, it follows that
\begin{equation}
\begin{split}
	& \frac{\partial \langle n_e \rangle}{\partial t}  
	+  n_{0,e}  \frac{\partial }{\partial x} \langle  \widetilde{b}_x  \widetilde{u}_{z,e} \rangle
	 =  0
\\
	\Rightarrow 
	&\frac{\partial \langle n_e \rangle}{\partial t}  
	- \frac{\partial }{\partial x} \frac{\langle   \widetilde{b}_x  \widetilde{J}_{z,e} \rangle}{|e|} 
	=  0,
	\label{eq: mean ne evolution}
\end{split}
\end{equation}
where $\widetilde{J}_{z,e} = -\widetilde{u}_{z,e}  n_{0,e} |e|$ is electron current density.
Note that the divergence of the electron current along tilted field lines (Amp\`ere's Law) is what determines $\langle n_e \rangle$ evolution.
Amp\`ere's Law states
\begin{equation}
	- \nabla_\perp^2 A_z = \mu_0 (J_{z,e} +J_{z,i}).
\end{equation}
Substitution into \Eref{eq: mean ne evolution} gives 
\begin{equation}
	\frac{\partial \langle n_e \rangle}{\partial t}  
	+ \frac{1}{\mu_0 |e|} \frac{\partial }{\partial x} \langle   \widetilde{b}_x  \nabla_\perp^2 \widetilde{A}_{z,e} \rangle
	+  n_{0,i}  \frac{\partial }{\partial x} \langle  \widetilde{b}_x  \widetilde{u}_{z,i} \rangle 
	=  0.
	\label{eq: n evol for Ji Je}
\end{equation}
In the last term on the RHS of \Eref{eq: n evol for Ji Je}, we take $u_{z} =u_{z,i}$ the parallel ion flow, consistent with our notation.
Using the magnetic Taylor identity \cite{Chen_2020}, we then obtain
\begin{equation}
	\frac{\partial \langle n_e \rangle}{\partial t}  
	+ \frac{\partial }{\partial x}   \Gamma_{e,s} =0,
\end{equation}
where 
\begin{equation}
	\Gamma_{e,s}  
	= \frac{-B_0}{\mu_0 |e|}  \frac{\partial }{\partial x} \langle \widetilde{b}_x \widetilde{b}_y \rangle
	+ n_0 \langle \widetilde{b}_x \widetilde{u}_z \rangle
\end{equation}
is the electron particle flux due to $\widetilde{b}$.
Note there are two contributions.
The first is the familiar piece due to the divergence of the Maxwell stress \cite{ding2008stochastic}. 
It arises from the flow of current along tilted field lines.
The second contribution $\langle \widetilde{b}_x \widetilde{u}_z \rangle$ studied here is due to \textit{ion} flow along tilted lines. 
Note both total and ion current contributions are required to calculate $\partial_t \langle n_e \rangle$. 
For the model analyzed here, \Eref{eq: bu caused by D_M} then gives
\begin{equation}
	\Gamma_{e,s}  
	= \frac{-B_0}{\mu_0 |e|} \frac{\partial }{\partial x} \langle \widetilde{b}_x \widetilde{b}_y \rangle
	- n_0 D_M \frac{\partial }{\partial x}  \langle u_z \rangle.
	\label{eq: RMP bump}
\end{equation}
The first term shows that stochastic lines and parallel ion flow gradient drive a net electron particle flux. 
The second piece adds to the familiar Maxwell force contribution. 
Note that this stochastic-field-induced particle flux likely is relevant to the phenomenon of `RMP pump-out.'
Since the rotation is necessary for stability with RMPs, the $\nabla \langle u_z \rangle$-driven flux contribution is of particular relevance. 
The discussion here clarifies the relations between \Eref{eq: mean evol. of n} and the electron particle flux.

\section{Calculating the Kinetic Stress and Compressive Energy Flux: Stochastic Fields and Turbulence} \label{sec: kinetic stress}
In \Sref{sec: Model}, we discussed the kinetic stress and compressive energy flux due to stochastic fields. 
In this section, we consider \textit{fluctuating} $E \times B$ flow effects.
These introduce a relatively fast scattering time scale that enters the response to $\widetilde{b}$.
We investigate the evolution of mean parallel flow and that of mean ion pressure (in presence of stochastic fields and turbulence) through the kinetic stress and compressive energy flux, respectively. 
Consider flow and pressure evolution in the basic model presented in \Sref{sec: Model}, we have the parallel acceleration and pressure equations:
\begin{equation}
	\frac{\partial}{\partial t} u_z + (\bold{u} \cdot \nabla) u_z = - \frac{1}{\rho} \nabla_z  p,
	\label{eq: parallel accel.}
\end{equation}
\begin{equation}
 \frac{\partial}{\partial t}p  + (\bold{u} \cdot \nabla) p = - \gamma p (\nabla_z \cdot \bold{u}_z ),
 \label{eq: pressure}
\end{equation}
where $z$ is set in parallel direction, and $x$ and $y$ are set in perpendicular direction. 
We decompose velocity and pressure as mean and its perturbation such that $\bold{u} = \langle \bold{u} \rangle + \widetilde{\bold{u}} $, $p = \langle p \rangle + \widetilde{p}$. 
By using mean field theory, we have
\begin{equation}
	\frac{\partial}{\partial t} \langle u_z \rangle  +  \frac{\partial}{\partial x} \langle \widetilde{u}_x \widetilde{u}_z \rangle  = - \frac{1}{\rho} \frac{\partial}{\partial x} \langle \widetilde{b}_x \widetilde{p} \rangle,
	\label{eq: toroidal equation}
\end{equation}
\begin{equation}
	\frac{\partial}{\partial t} \langle p \rangle 
	  +\frac{\partial}{\partial x}\langle \widetilde{u}_x \widetilde{p} \rangle 
	=-
	\rho c_s^2\frac{\partial}{\partial x}  \langle \widetilde{b}_x \widetilde{u}_z\rangle.
	\label{eq: pressure equation}
\end{equation}
The right hand side (RHS) of \Eref{eq: toroidal equation} is the divergence of the kinetic stress ($K$) such that
\[ 
-\frac{1}{\rho} \frac{\partial}{\partial x} \langle \widetilde{b}_x \widetilde{p}  \rangle \equiv  -\frac{\partial}{\partial x}K.
\]
The kinetic stress $K \equiv  \langle \widetilde{b}_x \widetilde{p} \rangle/\rho$ is determined by the stochastic magnetic field and the turbulence, as the pressure perturbation $\widetilde{p}$ is scattered by both the drift-wave turbulence and the stochastic field.
However, since it is the coherence of $\widetilde{b}_x$ and $\widetilde{p}$ that determines $K$, we seek $\widetilde{p} = (\delta p/\delta b) \widetilde{b}$, while including turbulent scattering in $\delta p/\delta b$.
Hence, the kinetic stress is derived by considering the $\widetilde{p}$ response to $\widetilde{b}_x$ that evolves in the presence of drift wave turbulence.
Notice that both the Reynolds stress $\langle \widetilde{u}_x \widetilde{u}_z\rangle$ and Kinetic stress $\langle \widetilde{p} \widetilde{b}_r \rangle/\rho $ in \Eref{eq: toroidal equation} are affected by stochastic magnetic fields. 
Chen et al. \cite{Chen_2021} discussed magnetic stochasticity effects on Reynolds stress.
Moreover, the RHS of \Eref{eq: pressure equation} contains the \textit{compressive energy flux} $H \equiv \rho c_s^2 \langle \widetilde{b}_x \widetilde{u}_z \rangle $, such that 
\[
-\rho c_s^2\frac{\partial}{\partial x}  \langle \widetilde{b}_x \widetilde{u}_z\rangle \equiv -\frac{\partial}{\partial x} H.
\]
This compressive energy flux $H$ describes the heat transport effect induced by compression along stochastic magnetic field lines.
This effect contributes to the evolution of mean pressure.
We do not elaborate further here on the electrostatic Reynolds stress $\langle  \widetilde{u}_x \widetilde{u}_z \rangle$ and the energy flux $\langle  \widetilde{u}_x \widetilde{p} \rangle$. 
Note for the former, correlation requires broken symmetry, the mechanisms for which are enumerated in Diamond et al. \cite{diamond2013overview}.
Note that details of the broken symmetry are not crucial to the remainder of this paper, so we ignore them hereafter.

We calculate the response of $\widetilde{p}$ and $\widetilde{u}_z$ to $\widetilde{b}_x$, so as to determine $K$ and $H$. 
However, we do so \textit{in the presence of scattering by drift-wave turbulence}.
Hence, \Eref{eq: parallel accel.} and \Eref{eq: pressure} yield
\begin{equation}
	-i \omega  \frac{\widetilde{p}}{\rho c_s} 
	+  c_s ik_z  \widetilde{u}_z 
	 + \widetilde{u}_{\perp} \nabla_\perp \frac{\widetilde{p}}{\rho c_s}
	= -\widetilde{u}_x \frac{\partial}{\partial x} \frac{\langle p \rangle}{\rho c_s}
	-\widetilde{b}_x c_s \frac{\partial}{\partial x} \langle u_z \rangle,
	\label{eq: pressure eq}
\end{equation}
\begin{equation}
	-i \omega  \widetilde{u}_z  +  c_s ik_z \frac{\widetilde{p}}{\rho c_s}   + \widetilde{u}_{\perp} \nabla_\perp \widetilde{u}_z
	= -\widetilde{u}_x \frac{\partial}{\partial x} \langle u_z \rangle
	-\widetilde{b}_x c_s \frac{\partial}{\partial x} (\frac{\langle p \rangle}{\rho c_s}).
	\label{eq: parallel acceleration}
\end{equation}
Note that the response to $\widetilde{u}_x$ on the RHS is not of interest since we take $\langle \widetilde{b}_x \widetilde{u}_x\rangle  = 0$,  i.e. we take drift waves and stochastic field uncorrelated, for simplicity. 
Here, the assumption $\langle \widetilde{b}_x \widetilde{u}_x\rangle  = 0 $ is based upon the assumed disparity in space-time scales, 
i.e. $\omega_{\widetilde{b}} \simeq 0 $ while $\omega_{\widetilde{u} } \simeq \omega_*$, and  $l_{\widetilde{b} } < l_{\widetilde{u}}$.
Thus, we take the drift wave turbulence as mesoscopic and quasi-Gaussian as usual, with statistics independent of the microscopic $\widetilde{b}_x$ (also taken as Gaussian).
Further detailed analysis of how non-zero correlation ($\langle \widetilde{b}_x \widetilde{u}_x\rangle \neq 0 $) might develop is given in Cao \& Diamond \cite{Cao_Diamond2021}. 
In particular, that paper shows the development of such correlation is a multi-scale effect and stems from maintaining $\nabla \cdot J =0$ on all scales. 
The details of this calculation are beyond the scope of this paper.  
By taking  \Eref{eq: parallel acceleration} $\pm$ \eref{eq: pressure eq}, we define the Riemann variables $f_{\pm,k\omega}\equiv \widetilde{u}_{z,k\omega} \pm \widetilde{p}_{k\omega}/\rho c_s$ and obtain 
\begin{equation}
	(-i\omega \pm ic_s k_z + i k_\perp \widetilde{u}_\perp ) f_{\pm,k\omega} 
	=
	-\widetilde{b}_x  c_s  
	\big(
	\frac{\partial}{\partial x} \frac{\langle p \rangle}{\rho c_s} \pm\frac{\partial}{\partial x}  \langle u_z \rangle 
	\big).
	\label{eq: Riemann variables}
\end{equation}
This is the evolution equation for the Riemann response to magnetic perturbation $\widetilde{b}_x$.
We compute the response of $\widetilde{u}_z$ and $\widetilde{p}$ to $\widetilde{b}_x$, which is \textit{static} --- i.e. has no time dependence.
And $\widetilde{u}_\perp$ is taken as stationary. 
Then, for $\omega \rightarrow 0$, we have:
\begin{equation}
	(\pm ic_s k_z + i k_\perp \widetilde{u}_\perp ) f_{\pm,k} 
	=
	-\widetilde{b}_x  c_s  
	\big(
	\frac{\partial}{\partial x} \frac{\langle p \rangle}{\rho c_s} \pm\frac{\partial}{\partial x}  \langle u_z \rangle 
	\big).
\end{equation}
Notice that the $\widetilde{u}_\perp \nabla_\perp$ operator in \Eref{eq: pressure eq} and \eref{eq: parallel acceleration} can be expressed as a cumulant scattering effect on a timescale long, compared with the auto-correlation time of drift-wave turbulence $\tau_{ac}$, i.e.
\begin{eqnarray}
 	\langle \frac{i}{-c_s k_z \mp \widetilde{u}_\perp k_\perp }  \rangle 
	&= \int d\tau e^{-ic_s k_z \tau} \langle  e^{\mp i \widetilde{u}_\perp k_\perp \int d\tau^{'} } \rangle
 	\\
	& = \int d\tau e^{-ic_s k_z \tau}   e^{-\underline{k}_i \underline{\underline{D}}_{ij,T} \underline{k}_j \tau} ,
\end{eqnarray}
where $
D_{ij,T} \equiv \sum\limits_k \widetilde{u}_{i,k} \widetilde{u}_{j,k} \tau_{ac}  
 $ is turbulent fluid diffusivity.
For perpendicular transport ($i=j=x $ or $y$), we have 
$
D_T \equiv \sum\limits_k |\widetilde{u}_{\perp,k}|^2 \tau_{ac} $, which generically is the order of the Gyro-Bohm diffusivity $D_{GB} \sim \rho_s^2 c_s /L_{n,\perp} 
$, as is $\nu_{turb}$.
Here, $L_{n,\perp} $ is density scale length. 
So, we can replace $\widetilde{u}_\perp \nabla_\perp$ with 
\begin{equation}
	\widetilde{u}_\perp \nabla_\perp 
	\equiv -\underline{\nabla}_\perp \cdot \underline{\underline{D_T}}\cdot \underline{\nabla}_\perp.
\end{equation}
Hence, \Eref{eq: Riemann variables} become
\begin{equation}
	(\pm ic_s k_z +  k_\perp^2 D_T ) f_{\pm,k} 
	=
	-\widetilde{b}_x  c_s  
	\big(
	\frac{\partial}{\partial x} \frac{\langle p \rangle}{\rho c_s} \pm\frac{\partial}{\partial x}  \langle u_z \rangle 
	\big).
\end{equation}  
From this equation, we have 
\begin{eqnarray}
\begin{split}
	\frac{\widetilde{p}_k }{\rho c_s}   
	&= \frac{1}{2}(f_{+,k} -f_{-,k})
	= \frac{-1}{k_\perp^4 D_T^2 + k_z^2 c_s^2} \times
	\\
	&\bigg[ \widetilde{b}_{x,k}c_s k_\perp^2 D_T \frac{\partial}{\partial x} \langle u_z \rangle - i k_z c_s^2 \widetilde{b}_{x,k}\frac{\partial}{\partial x} (\frac{\langle p \rangle}{\rho c_s})
	\bigg],
	\label{eq: p response with b}
\end{split}
	\\
\begin{split}
	\widetilde{u}_{z,k}    
	&= \frac{1}{2}(f_{+,k} +f_{-,k})
	= \frac{-1}{k_\perp^4 D_T^2 + k_z^2 c_s^2} \times
	\\
	&\bigg[ \widetilde{b}_{x,k}c_s k_\perp^2 D_T \frac{\partial}{\partial x}  (\frac{\langle p \rangle}{\rho c_s}) - i k_z c_s^2 \widetilde{b}_{x,k}\frac{\partial}{\partial x}  \langle u_z \rangle
	\bigg].
	\label{eq: uz response with b}
\end{split}
\end{eqnarray}
Then, \Eref{eq: p response with b} and \Eref{eq: uz response with b} yield
\begin{equation}
\begin{split}
	K = \frac{1}{\rho} \langle \widetilde{b}_x \widetilde{p} \rangle 
	= \frac{1}{\rho} \sum\limits_{k_y,k_z} |\widetilde{b}_{x,k}|^2 
	\frac{-1}{k_\perp^4 D_T^2 + k_z^2 c_s^2} 
	\\
	\bigg[ \rho c_s^2 k_\perp^2 D_T \frac{\partial}{\partial x} \langle u_z \rangle  
	- i k_z c_s^2  \frac{\partial}{\partial x} \langle p \rangle
	\bigg].
	\label{eq: bp_1}
\end{split}
\end{equation}
\begin{equation}
\begin{split}
	H \equiv \rho c_s^2 \langle \widetilde{b}_x \widetilde{u}_z \rangle 
	=  \rho c_s^2 \sum\limits_{k_y,k_z} |\widetilde{b}_{x,k}|^2 \frac{-1}{k_\perp^4 D_T^2 + k_z^2 c_s^2} 
	\\
	\bigg[   k_\perp^2 D_T \frac{\partial}{\partial x} \frac{  \langle p \rangle}{\rho}
	- i k_z  c_s^2 \frac{\partial}{\partial x}  \langle u_z \rangle 
	\bigg].
	\label{eq: bu_1}
\end{split}
\end{equation}
The  denominator of the response function $1/(k_\perp^4 D_T^2 + k_z^2 c_s^2)$ can be approximated as 
$
	(k_\perp^2 D_T)^2 \simeq 1/\tau_{c,k}^2,
$
where $\tau_{c} $ is the decorrelation time due to perpendicular turbulent scattering. 
The significance of the factor of $i$ in the second term of the source will be apparent when considering reduction to the quasilinear limit (see \Sref{sec: Weak Turb}).
Non-zero correlations $ \langle \widetilde{b}_x \widetilde{u}_z \rangle $ and $ \langle \widetilde{b}_x \widetilde{p} \rangle$, which contribute the kinetic stress $K$ and compressive energy flux $H$, are due to the synergetic effect of the perpendicular turbulent mixing ($k_\perp^2 D_T$) and stochastic magnetic field ($|\widetilde{b}_x|^2$) scattering, via gradients of mean parallel flow $\partial_x \langle u_z \rangle $ and mean pressure $\partial_x \langle p \rangle$. 
Also, by observing the denominator of the responses, one can notice that $K$ and $H$ can be set by different mechanisms.
When $k_\perp^2 D_T > k_z c_s $, the decorrelation due to scattering is stronger than that due to acoustic signal decoherence.
For $ k_z c_s > k_\perp^2 D_T$, we recover the quasilinear results.
These two regimes will be discussed further in \Sref{sec: Strong turb} and \sref{sec: Weak Turb}.
Finally, note that for $k_\perp^2 D_T > k_\parallel c_s$, the determinant of the transport matrix is positive, so the stability of transport equations (with $K$ and $H$ only) is assured. 
\subsection{Calculating the flux} \label{sec: pre K and H}
In the following \Sref{sec: pre K and H}, \ref{sec: Strong turb}, and \ref{sec: Weak Turb}, we consider the effect of magnetic shear in presence of stochastic fields.
We shall calculate the kinetic stress and compressive energy flux in detail. 
Sheared magnetic field geometry is used to clarify aspects of the competition between acoustic pulse decorrelation at rate $k_z c_s$ and turbulent scattering,  with rate $k_\perp^2 D_T$, and its implication for the structure of the fluxes. 
Attention here is focused on the interplay of different terms in the expressions for $K$ and $H$.

The second term in the denominator of the response function in \Eref{eq: bp_1} and \eref{eq: bu_1} can be approximated as 
$
	k_z^2 c_s^2  
	= (k_y x /L_s)^2 c_s^2,
$
where $L_s$ the is magnetic shear length such that
$1/L_s =  q^\prime r_0/q^2 R_0$, $q^\prime \equiv dq/dx$, $q$ is the safety factor, and $x$ is the \textit{distance from the resonant surface of the perturbation} --- i.e. $x = r -r_{m,n}$, where $m$, $n$ are the poloidal and toroidal mode numbers, respectively.
We decompose \Eref{eq: bp_1} into two parts 
\begin{equation}
\begin{split}
	\langle \widetilde{b}_x \widetilde{p} \rangle 
	&=  \underbrace{\sum\limits_{k_y,k_z}  |\widetilde{b}_{x,k}|^2 \frac{\tau_{c,k}}{1 + (k_y x/L_s)^2 c_s^2 \tau_{c,k}^2} 
	\bigg( -\rho  c_s^2  \frac{\partial}{\partial x} \langle u_z \rangle  \bigg)
	}_{\textcircled{a}}
	\\
	&+ \underbrace{\sum\limits_{k_y,k_z} |\widetilde{b}_{x,k}|^2 \frac{1}{(k_\perp^2 D_T)^2 + k_z^2 c_s^2}  
	\bigg( i k_z c_s^2  \frac{\partial}{\partial x} \langle p \rangle 
	\bigg)
	}_{\textcircled{b}}.
	\label{eq: bp_2}
\end{split}
\end{equation}

\begin{figure}[h!]
\centering
  \includegraphics[scale=0.2]{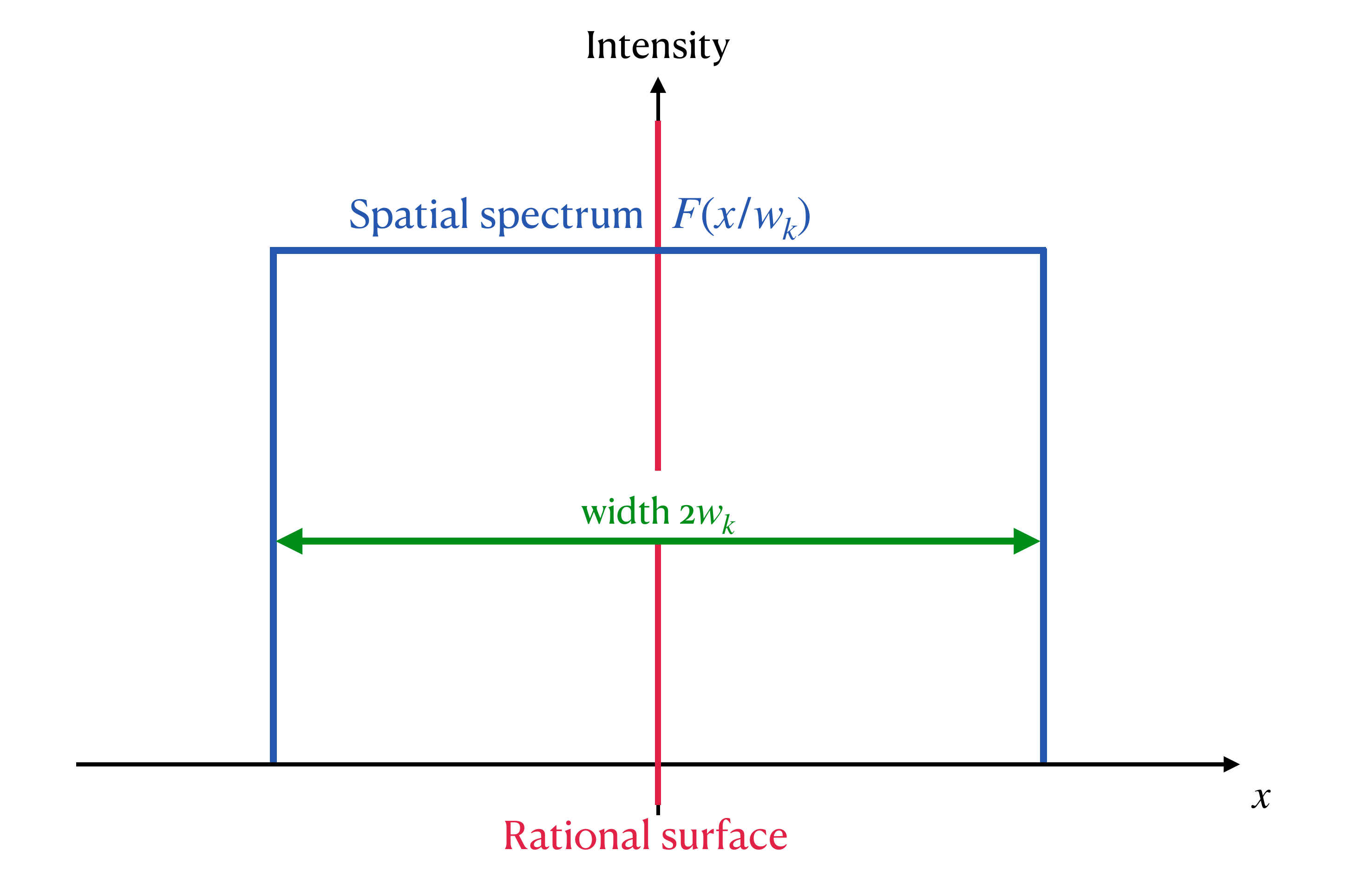}
  \caption{$ F(x/w_k)$ is spatial spectrum for $|\widetilde{b}_{x,k}|^2$ in radial direction. Here we define $x = r-r_{m,n}$, where $r_{m,n}$ is the location of a rational surface with mode number $m, \; n$.  }
  \label{fig:Material_7}
\end{figure}

\begin{figure}[h!]
\centering
  \includegraphics[scale=0.15]{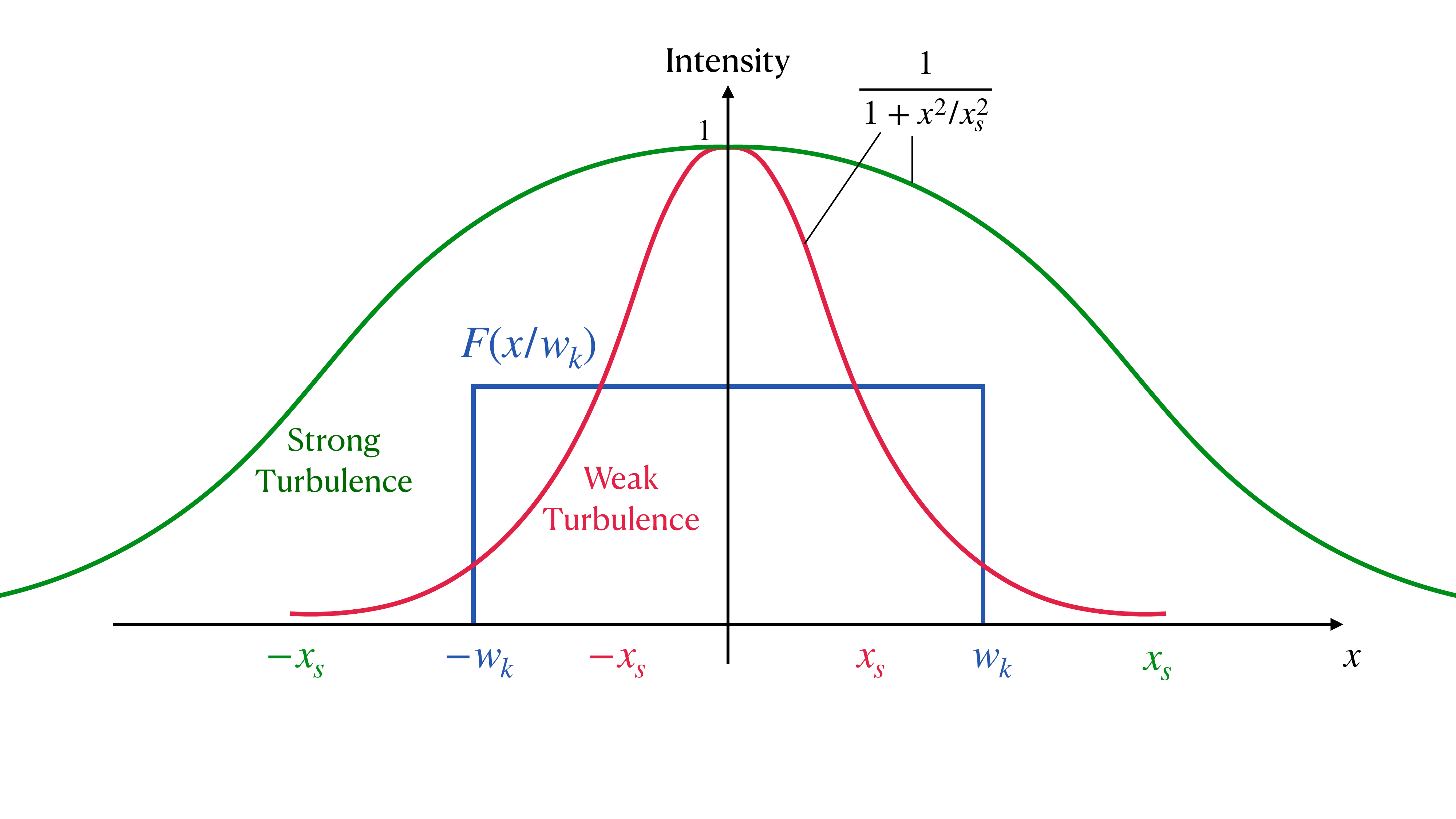}
  \caption{ The integral of spatial spectrum of stochastic field and the response function $\int dx F(x/w_k)
	 \cdot
	 [1/1 + (x/x_s)^2]$.
  Green and red lines indicate response functions in strong and weak turbulence regime, respectively.
  For strong turbulence ($x_s \gg w_k$), $w_k$ is the cutoff length in the integral. 
For weak turbulence ($x_s \ll w_k$), $x_s$ is the cutoff length scale.}
  \label{fig:Material_8}
\end{figure}

The spectral sum relevant to this shear effect is $\sum\limits_{k_y,k_z}$.
The radial structure is accounted for by the introduction of a box function $F(x/w_k)$ for the magnetic perturbation intensity, which we further analyze in the following paragraph. 
We approximate the discrete summation $\sum \limits_{k_yk_z}$ by a continuous integral:
\begin{equation}
	\sum  \limits_{k_yk_z}= \sum\limits_{m,n}
	 = \int dm \int dn.
\end{equation}
By using $n = m/q$ and $dn = |m| q^\prime dx/q^2$, we have
\begin{equation}
	\int dm \int dn =  \int dm \int dx \frac{|m|}{q^2} q^\prime
	= r_0 \int dk_y \int dx \frac{|k_y|}{q} \hat{s},
\end{equation}
where $\hat{s} = \frac{ r_0}{q} \frac{ dq}{dr}$ is the magnetic shear. 
Now, we write the magnetic perturbation spectrum $|\widetilde{b}_{x,k}|^2$ as 
\[ |\widetilde{b}_{x,k}|^2 = C S(k_y) F(x/w_k),
\]
where $C$ is a normalization constant, $S(k_y)$ is the k-spectrum of $\widetilde{b}_x$, $F(x/w_k)$ is the spatial spectrum form factor, and $w_k$ is the spatial width of $|\widetilde{b}_{x,k}|^2$ (see figure \ref{fig:Material_7}) .
We assume $|\widetilde{b}_{x,k}|^2$ perturbations are densely packed and take the spatial spectrum $F(x/w_k)$ to be a normalized box function such that $\int dx F(x/w_k) =1$.
Hereafter, we define the intensity of magnetic perturbation $b_{x,0}^2$ as
\begin{eqnarray}
	b_{x,0}^2 \equiv \sum  \limits_{k_yk_z} |\widetilde{b}_{x,k}|^2 
	&= r_0\int dk_y   \int dx \frac{ k_y \hat{s}}{q} \cdot C S(k_y) F(x/w_k),
	 	 \nonumber
	\\
	&= C \int dk_y  \frac{k_y r_0^2 q^\prime}{q^2}  S(k_y) \underbrace{\int dx F(x/w_k)}_{=1},
	\nonumber
	\\
	 &= C \int dk_y  \frac{k_y r_0^2 q^\prime}{q^2}  S(k_y) .
\end{eqnarray}
The normalization constant $C$ hence is defined as
\begin{equation}
	 C \equiv \frac{b_{x,0}^2}{\int dk_y \frac{k_y r_0^2 q^\prime}{q^2}  S(k_y) }
\end{equation}
where $m/r_0 \equiv k_y$, $R_0$ and $r_0$ are the major and minor radius, respectively.
The first term in equation (\ref{eq: bp_2}) becomes
\begin{equation}
\begin{split}
	\textcircled{a}
	 &=\sum  \limits_{k_yk_z} |\widetilde{b}_{x,k}|^2 \frac{\tau_{c,k}}{1 + (k_y x/L_s)^2 c_s^2 \tau_{c,k}^2} 
	\bigg( -\rho  c_s^2  \frac{\partial}{\partial x} \langle u_z \rangle  \bigg)
	 \\
	 &= 
	 \underbrace{
	 C \int dk_y  \frac{k_y r_0^2 q^\prime}{q^2}  S(k_y) \int dx F(x/w_k)
	 }_{\sum  \limits_{k_yk_z} |\widetilde{b}_{x,k}|^2} 
	 \cdot
	 \\
	& \;\;\;
	\frac{\tau_{c,k}}{1 + (k_y x/L_s)^2 c_s^2 \tau_{c,k}^2} 
	\bigg( -\rho  c_s^2  \frac{\partial}{\partial x} \langle u_z \rangle  \bigg)
	\label{eq: circle a}
\end{split}
\end{equation}
The response function $ 1/(1 + (k_y x/L_s c_s \tau_{c,k})^2)$ in the equation is the key to understanding the physics of pressure evolution.
We define the \textbf{acoustic width} ($x_s$) by 
\begin{equation}
	x_s \equiv \frac{L_s}{k_y c_s \tau_{c,k}},
\end{equation}
The acoustic width is the value of $x$ for which $k_z c_s(x) =1/\tau_{c,k}(x)$, where $k_z = k_z(x)$.
So, $x_s =L_s/k_y c_s \tau_{c,k}$.
Loosely speaking, $x_s$ defines the location where the rate of parallel acoustic streaming equals the decorrelation rate.
Here $x_s$ is analogous to the familiar $x_i = \omega L_s/  k_y v_{th,i}$ ---  the ion Landau resonance point, where $v_{th,ix}$ is the ion thermal speed \cite{landau1946vibrations}. 
The $\tau_{c,k}$ sets the acoustic width --- e.g. in strong fluid turbulence (small $\tau_{c}$), $x_s$ is large; in weak fluid turbulence, $x_s$ is small.
Hence, the first term of \Eref{eq: bp_2} becomes
\begin{equation}
\begin{split}
	\textcircled{a}
	 =
	  C \int dk_y  \frac{k_y r_0^2 q^\prime}{q^2}  S(k_y) 
	  \int dx F(x/w_k)
	  \cdot
	  \\
	 \frac{\tau_{c,k}}{1 + (x/x_s)^2} 
	\big( -\rho  c_s^2  \frac{\partial}{\partial x} \langle u_z \rangle  \big)
	\label{eq: circled a_2 }
\end{split}
\end{equation}
For strong turbulence, $\tau_{c,k}$ is small, such that $x_s \gg w_k$.
So $w_k$ is the cutoff length in the integral (see \Fref{fig:Material_8}, green curve). 
For weak turbulence (i.e. $x_s \ll w_k$), however, the acoustic width $x_s$ is the cutoff length scale (see \fref{fig:Material_8}, red curve).
Let's consider these two limits. 
\subsection{Strong Turbulence} \label{sec: Strong turb}
In strong fluid turbulence, we have $x_s \gg w_k$ ( or $k_\perp^2 D_T > k_z c_s$). 
Recall in \Eref{eq: bp_2}, $\langle \widetilde{b}_x \widetilde{p}  \rangle = \textcircled{a}+\textcircled{b}$.
Here, the integral $\int d(x)
	 \frac{  F(x/w_k)}{1 + \cancel{x^2/x_s^2}}  \simeq 1$.
So, the first term in \Eref{eq: bp_2} becomes
\begin{equation}
	\textcircled{a}
	 \simeq -\rho  c_s^2 \sum  \limits_{k_yk_z} |\widetilde{b}_{x,k}|^2 \tau_{c,k}  \frac{\partial}{\partial x} \langle u_z \rangle.
	\label{eq: circled a_strong}
\end{equation}
The second term in \Eref{eq: bp_2}, assuming $k_\perp^4 D_T^2 \gg k_z^2 c_s^2$, becomes small
\begin{equation}
	\textcircled{b} 
	= \sum  \limits_{k_yk_z} |\widetilde{b}_{x,k}|^2 \frac{1}{(k_\perp^2 D_T)^2 + k_z^2 c_s^2}  
	\bigg( i k_z c_s^2  \frac{\partial}{\partial x} \langle p \rangle 
	\bigg)
	\rightarrow 0,
\end{equation}
in the limit $ k_\perp^4 D_T^2 \gg k_z^2 c_s^2 $.
Derivation details can be found in \ref{appen: Strong turb}.
Hence, the kinetic stress can be simplified to 
\begin{equation}
	\langle \widetilde{b}_x \widetilde{p} \rangle 
	\simeq -\rho  c_s^2 \sum  \limits_{k_yk_z} |\widetilde{b}_{x,k}|^2 \tau_{c,k}  \frac{\partial}{\partial x} \langle u_z \rangle.
\end{equation}
This indicates that \textit{in the presence of strong scattering, the kinetic stress depends on the electrostatic $\tau_{c,k}$}.
The kinetic stress hence becomes simply:
\begin{equation}
	 K \equiv 
	  \frac{1}{\rho} \langle \widetilde{b}_x \widetilde{p} \rangle 
	\simeq - \sum  \limits_{k_yk_z} |\widetilde{b}_{x,k}|^2 \tau_{c,k} c_s^2  \frac{\partial}{\partial x} \langle u_z \rangle.
\end{equation}
From this, we recover a \textit{hybrid viscosity} produced by stochastic magnetic fields $|\widetilde{b}_{x,k}|^2 c_s^2 $, with a correlation time set by electrostatic scattering $\tau_{c,k}$.
Hence,
\begin{equation}
\begin{split}
	D_{st}(x)
	&\equiv  
	\frac{b_{x,0}^2 \int dk_y  \frac{k_y r_0^2 q^\prime}{q^2}   \tau_{c,k}  S(k_y) c_s^2}{\int dk_y  \frac{k_y r_0^2 q^\prime}{q^2}   S(k_y)  }	
\\
	& \simeq 
	\sum  \limits_{k_yk_z} |\widetilde{b}_{x,k}|^2 (x)\frac{c_s^2 }{k_\perp^2 D_T},
	\label{eq: turb. viscosity}
\end{split}
\end{equation}
\Eref{eq: turb. viscosity} leads us to notice that the combined effects of stochastic fields in the numerator and (electrostatic) turbulent scattering in the denominator \textit{together}  define $D_{st}$. 
This hybrid turbulent viscosity is the actual diffusivity that describes how the mean flow is scattered by stochastic magnetic fields.
The parallel flow evolution equation then becomes
\begin{equation}
	\frac{\partial}{\partial t} \langle u_z \rangle 
	 = -\frac{\partial}{\partial x}\langle \widetilde{u}_x \widetilde{u}_z \rangle 
	+
	\frac{\partial}{\partial x} D_{st}(x) \frac{\partial}{\partial x} \langle u_z \rangle.
\end{equation}
\textit{This indicate that the turbulent viscous stress balances $\widetilde{b}_x \partial_x \langle u_z \rangle$}.


Similarly,  \Eref{eq: bu_1} gives the compressive energy flux (H)
\begin{equation}
\begin{split}
	H (x)
	&\equiv \rho c_s^2 \langle \widetilde{b}_x \widetilde{u}_z\rangle
	\\
	&\simeq - c_s^2 \sum  \limits_{k_yk_z} |\widetilde{b}_{x,k}|^2 \tau_{c,k}  \frac{\partial}{\partial x} \langle p \rangle
	\simeq - D_{st} (x) \frac{\partial}{\partial x} \langle p \rangle.
	\label{eq: H in strong turb.}
\end{split}
\end{equation}
This indicates that \textit{the tilting of the magnetic field lines in presence of the pressure gradient (i.e. $\widetilde{b}_x\partial \langle p \rangle/\partial x \neq 0$) balances the turbulent diffusion}. 
Notice that \Eref{eq: H in strong turb.} also shows that $D_T \nabla_\perp^2 \widetilde{u}_z \simeq - \sum  \limits_{k_yk_z} \widetilde{b}_{x,k} \partial_x (\langle p\rangle / \rho)$.
This again indicates that the \textit{change in mean pressure ($\partial_x \langle p \rangle/ \rho $) due to the stochastic fields is balanced by turbulent mixing of parallel flow} ($\nabla_\perp^2 \widetilde{ u}_z $, see \Fref{fig:Weak_turb_pressure}).
The pressure equation now can be written as
\begin{equation}
	\frac{\partial}{\partial t} \langle p \rangle 
	 = -\frac{\partial}{\partial x}\langle \widetilde{u}_x \widetilde{p} \rangle 
	+
	\frac{\partial}{\partial x} D_{st}  (x)\frac{\partial}{\partial x} \langle p \rangle,
\end{equation}
again a diffusion equation.

\subsection{Weak Turbulence} \label{sec: Weak Turb}
For weak fluid turbulence, we have $ w_k \gg x_s$ ( or $k_z c_s > k_\perp^2 D_T  $). 
Recall \Eref{eq: circled a_2 }
\begin{equation}
\begin{split}
	\textcircled{a}
	 =
	 & C \int dk_y  \frac{k_y r_0^2 q^\prime}{q^2}  S(k_y) \int dx F(x/w_k)
	 \cdot
	 \\
	& \frac{\tau_{c,k}}{1 + (x/x_s)^2}  \cdot \bigg( -\rho  c_s^2  \frac{\partial}{\partial x} \langle u_z \rangle  \bigg).
\end{split}
\end{equation}
Since in the weak turbulence limit, the cutoff of integral is set by $x_s$ and hence $F(x/w_k) \simeq F(x_s/w_k\rightarrow 0)$.
So, $\textcircled{a}$ is simplified as follows:
\begin{equation}
\begin{split}
	\textcircled{a}
	&\simeq 
	\frac{b_{x,0}^2 \cancel{\int dk_y  \frac{k_y r_0^2 q^\prime}{q^2}  S(k_y)}}{\cancel{\int dk_y \frac{k_y r_0^2 q^\prime}{q^2}  S(k_y) }}
	\cdot 
	 \tau_{c,k} \frac{x_s}{w_k} 
	F(0)
	 \cdot
	 \\
	& \;\;\;
	\underbrace{\int\limits^{x =x_s}_0 d(x/x_s)
	 \frac{  1}{1 + (x/x_s)^2} }_{=arctan(x_s/x_s) = \pi/4}
	\bigg( -\rho  c_s^2  \frac{\partial}{\partial x} \langle u_z \rangle  \bigg)
	\\
	& \simeq
	-b_{x,0}^2 \tau_{d,k} \frac{x_s}{w_k}  F(0)\frac{\pi}{4}\bigg( \rho  c_s^2  \frac{\partial}{\partial x} \langle u_z \rangle  \bigg) 
\end{split}
\end{equation}
where $\tau_{d,k}\equiv   L_s /k_y c_s w_k$ is dispersal timescale of an acoustic wave packet propagating along the stochastic magnetic field. 
This dispersal timescale $\tau_{d,k}$ defines the width of the acoustic signal `cone'.
The second term in \Eref{eq: bp_2} becomes
\begin{equation}
\begin{split}
	\textcircled{b} 
	&= \sum  \limits_{k_yk_z} |\widetilde{b}_{x,k}|^2 \frac{1}{\cancel{(k_\perp^2 D_T)^2} + k_z^2 c_s^2}  
	\big( i k_z c_s^2  \frac{\partial}{\partial x} \langle p \rangle 
	\big),
	   \\
	   & \simeq-D_M \frac{\partial}{\partial x} \langle p \rangle,
\end{split}
\end{equation}  
where $D_M(x) \equiv \sum  \limits_{k_yk_z} |\widetilde{b}_{x,k}|^2 \tau_{d,k}(x) c_s$ is the magnetic diffusivity.
Hence, the kinetic stress flux is 
\begin{equation}
	\langle \widetilde{b}_x \widetilde{p} \rangle 
	=
	-b_{x,0}^2 \tau_{d,k} F(0)\frac{\pi}{4} \rho  c_s^2  \frac{\partial}{\partial x} \langle u_z \rangle 
	 -D_M	  \frac{\partial}{\partial x} \langle p \rangle .
	\label{eq: p perturb. in weak turb.}
\end{equation}
The first term on the RHS is asymptotically small, so $\textcircled{a} \rightarrow 0$ for $x_s/w_k  \rightarrow 0$ in this limit.
The detailed calculation is shown in \ref{appen:Weak trub}. 
Notice that \Eref{eq: p perturb. in weak turb.} also shows that $\nabla_z \widetilde{p} \simeq -\sum\limits_k \widetilde{b}_{x,k} \partial_x \langle p\rangle$, by approximating $1/\tau_{d,k} c_s $ with operator $\nabla_z$.
This indicates that the \textit{change in mean pressure ($\partial_x \langle p \rangle $) due to the stochastic fields is balanced by a parallel pressure gradient} ($\nabla_z \widetilde{ p} $, see \Fref{fig:Weak_turb_pressure}).
The kinetic stress in this limit can be simplified as
\begin{equation}
	 K(x) 
	\equiv  \frac{1}{\rho}  \langle \widetilde{b}_x \widetilde{p} \rangle 
	\simeq 
	 -\frac{1}{\rho}	D_M(x) \frac{\partial}{\partial x} \langle p \rangle .
	 \label{eq: weak K result}
\end{equation}
Hence, we have the parallel flow evolution
\begin{equation}
	\frac{\partial}{\partial t} \langle u_z \rangle 
	 \simeq -\frac{\partial}{\partial x}\langle \widetilde{u}_x \widetilde{u}_z \rangle 
	 +  \frac{\partial}{\partial x}\frac{D_M(x)}{\rho}	    \frac{\partial}{\partial x} \langle p \rangle .
\end{equation}

Similarly, we have the compressive energy flux 
\begin{equation}
	H (x)
	= \rho c_s^2 \langle \widetilde{b}_x \widetilde{u}_z \rangle
	= -\rho c_s^2 D_M (x) \frac{\partial}{\partial x} \langle u_z \rangle.
	\label{sec: weak H result}
\end{equation}
Notice that this equation shows the response of mean parallel flow, due to stochastic field tilting ($\widetilde{b}_x \partial \langle  u_z \rangle /\partial x $), is balanced by the parallel flow compression ($\nabla_\parallel \widetilde{u}_z $), i.e. equivalent to $\underline{B_z} \cdot \underline{\nabla}  u_z =0$ or $\nabla_\parallel \widetilde{u}_z = -\widetilde{b}_x \partial \langle  u_z \rangle /\partial x $.
Hence the pressure equation can be written as
\begin{equation}
	\frac{\partial}{\partial t} \langle p \rangle 
	 = -\frac{\partial}{\partial x}\langle \widetilde{u}_x \widetilde{p} \rangle 
	+
	\frac{\partial}{\partial x}\rho c_s^2 D_M(x) \frac{\partial}{\partial x} \langle u_z \rangle.
\end{equation}

\Eref{eq: weak K result} and \eref{sec: weak H result} indicate that for weak scattering, momentum and energy transport occur only through stochastic fields, with the familiar transport coefficient $c_s D_M$.
There is no dependence on $D_T$ for $k_\perp^2 D_T \ll k_z c_s$.
This result is equivalent to that in FGC \cite{finn1992}. 
Note, however, that the key effect for $\langle u_z \rangle$ is residual stress; and for $\langle p \rangle $, it is an off-diagonal flux. 
The comparison of $K$ and $H$ in strong and weak turbulence regime is shown in \Tref{table}.

\section{Discussion} \label{sec: Discuss}
In this paper, we develop the theory of ion heat and parallel momentum transport due to stochastic magnetic fields and turbulence. 
We focus on the kinetic stress ($K$) and the compressional flux ($H$) due to stochastic fields in the presence of (electrostatic) turbulence. 
The responses $\delta p/\delta b$ and $\delta u_\parallel/\delta b$ are calculated by integration over \textit{perturbed} particle trajectories and then used to close the fluxes.  
Interestingly, this analysis renders moot one of the deepest questions in stochastic-field-induced transport. 
Recall that Rechester and Rosenbluth\cite{Rosenbluth1978} showed that irreversibility requires some means to scatter particles off magnetic field lines, lest they bounce back and undergo no net excursion. 
Here, ambient cross-field electrostatic scattering supplies this needed effect. 
Thus, $\delta p/\delta b$ and $\delta u_\parallel/\delta b$ should be viewed as statistically averaged nonlinear responses. 
Here, we posit an ambient ensemble of drift waves, which specifies $\langle \widetilde{u}_\perp^2 \rangle$.
The probability distribution functions (PDFs) of $\langle \widetilde{u}_\perp^2 \rangle$ and $\langle \widetilde{b}_x^2 \rangle$ are assumed to be quasi-Gaussian and independent. 
General results are obtained and shown to recover the dynamic balance limit (viscous dissipation vs. $\widetilde{b}_x \partial \langle p \rangle/\partial x$, for $k_\perp^2 D_T > k_z c_s$) and the parallel pressure balance limit ($\nabla_z \widetilde{ p} $ vs. $\widetilde{b}_x \partial \langle p \rangle/\partial x$, for $k_z c_s > k_\perp^2 D_T  $), as appropriate. 
In reality, dynamic balance is the relevant case, and the quasilinear regime is of very limited practical interest. 
We calculate the explicit form of the turbulent viscous flux and compressive energy flux, and show both are diffusive with a hybrid diffusivity $D_{st} \equiv \sum\limits_k |\widetilde{b}_{x,k}^2| c_s^2/k_\perp^2 D_T$ --- i.e. determined by magnetic fluctuations but with correlation time set by turbulent scattering. 
The hybrid diffusivity $D_{st}$ is sensitive to the long wavelength content of $|\widetilde{b}_k^2|$.
The analysis is extended to the case of a sheared mean magnetic field. 
We show that the critical comparison is between the $|\widetilde{b}|^2$ spatial spectral width ($w_k$) and the acoustic width, i.e. $x_s = L_s/k_y c_s \tau_{c,k}$, where $\tau_{c,k} $ is decorrelation time due to perpendicular turbulent scattering.

This paper explores relatively untouched issues, namely the interaction of stochastic magnetic field and turbulence, and how they together drive transport.
As such, several of the results merit further discussion.
First, while the analysis is in the spirit of a resonance broadening calculation, the basic \textit{form} of the flux-gradient relation charges with the ratio of $k_z c_s$ to $k_\perp^2 D_T$.
Indeed, the kinetic stress changes from a residual stress to a turbulence viscous stress.
Also, given that $k_\perp^2 D_T \simeq \omega \gg k_z c_s$, the strong turbulence regime results are surely the relevant ones, and it is unlikely that the pure quasilinear predictions are ever observed.
This point is the major \textit{prediction} of this paper. 
This outcome is in contrast to the case for the quasilinear predictions for electromagnetic turbulence \cite{Peng_2016}, which are more robust since $\omega$, not $k_z c_s$, is the relevant base rate there.
Second, the sensitivity of the hybrid diffusivity $D_{st} \equiv \sum\limits_k |\widetilde{b}_{x,k}^2| c_s^2/k_\perp^2 D_T$ to long wavelength (i.e. `slow modes') is interesting and reminiscent of the results of Taylor and McNamara \cite{taylor1971plasma}.
Further study, including coupling to $E\times B$ shearing, is needed. 

Results of this paper are amenable to testing via computer simulations. 
Such studies would necessarily be non-trivial, as they require simulation of turbulence in stochastic fields. 
Studies might compare the kinetic stress and compressive energy flux calculated directly from the simulation to the predictions made here. 
Turbulence intensity could be scanned by varying the deviation from marginality.
In this way, one should be able to pass smoothly from the weak turbulence/quasilinear regime ($k_\perp^2 D_T\ll k_z c_s$) to the strong turbulence/nonlinear regime ($k_\perp^2 D_T\gg k_z c_s$), and evaluate scaling trends in both limits. 

Several questions and extensions for further study naturally suggest themselves. 
Magnetic drifts could be included in theory, which could then be used to study the effect of stochastic magnetic fields and turbulence upon geodesic acoustic modes (GAMs) \cite{Winsor1968, Itoh_2005, Melnikov_2006, Miki2010,zarzoso2013}. 
This topic is of obvious relevance to edge turbulence and transitions. 
Second, we have assumed throughout the magnetic perturbations and electrostatic turbulence are uncorrelated, i.e. $\langle \widetilde{b}_x \widetilde{\phi}  \rangle=0$.
Recent results, however, indicate that the constraint of $\nabla \cdot J =0$ naturally forces the generation of small scale convective cells by the interaction of long wavelength flows with turbulence.
As a consequence, a non-zero $\langle \widetilde{b}_x \widetilde{\phi}  \rangle$ develops, indicative of small-scale correlations between turbulence and stochastic fields. 
These may induce novel cross-coupling in the fluxes. 
Work on this question is ongoing. 
Finally, since the system studied here essentially is one of gas dynamics in a stochastic field, we note it may have relevance to problems in cosmic ray acceleration and propagation \cite{malkov2009nonlinear}.
In those problems, magnetic irregularities are thought to be scatter particles in turbulent environments --- similar to the physics discussed in this paper.   
\ack{
We thank Mingyun Cao, Lu Wang, Weixin Guo, TS Hahm, and Xavier Garbet for helpful discussions. We also acknowledge stimulating interactions with participants of the 2021 Festival de Th\'eorie and the 2021 KITP program Staircase 21. KITP is supported in part by the National Science Foundation under Grant No. NSF PHY-1748958. This research was supported by the U.S. Department of Energy, Office of Science, Office of Fusion Energy Sciences, under Award No. DE-FG02–04ER54738.  SMT would like to acknowledge support of funding from
the European Research Council (ERC) under the European
Union’s Horizon2020 research and innovation programme
(grant agreement no. D5S-DLV-786780)}

\begin{table}
\caption{\label{tabone}
A comparison of strong and weak turbulent MHD for Kinetic stress and compressive energy flux. 
} 

\begin{indented}
\lineup
\item[]\begin{tabular}{@{}*{7}{l}}
\br                              
& Strong Turbulence & Weak Turbulence\cr 
\mr
\0\0  Kinetic Stress
	\cr 
\0\0 $K \equiv \langle \widetilde{b}_x \widetilde{p}   \rangle /\rho$ 
&$K = - D_{st} \frac{\partial}{\partial x} \langle u_z \rangle$  
& $ K  = -D_M \frac{\partial}{\partial x} \langle p \rangle$  
\cr
\mr
 Compressive energy flux
	\cr
\0\0 $H \equiv \rho c_s^2 \langle \widetilde{b}_x \widetilde{u}_z\rangle$
 &  $H = -  D_{st} \frac{\partial}{\partial x} \langle p \rangle$ 
& $H =  -\rho c_s^2 D_M \frac{\partial}{\partial x} \langle u_z \rangle$
\cr 
\br
\end{tabular}
\end{indented}
	\label{table}
\end{table}
\section{DATA AVAILABILITY}
The data that support the findings of this study are available from the corresponding author upon reasonable request.

\onecolumn
\appendix
\section{Strong Turbulence Limit} \label{appen: Strong turb} 
In strong fluid turbulence, we have $x_s > w_k$ (or $k_\perp^2 D_T> k_\parallel c_s$)---$w_k$ sets a cut-off for the integral $\int dx$.
The first term in \Eref{eq: bp_2} becomes
\begin{equation}
\begin{split}
	\textcircled{a}
	 &=\sum\limits_{k_y k_z} |\widetilde{b}_{x,k}|^2 \frac{\tau_{c,k}}{1 + (k_y x/L_s)^2 c_s^2 \tau_{c,k}^2} 
	\bigg( -\rho  c_s^2  \frac{\partial}{\partial x} \langle u_z \rangle  \bigg)
	\\
	& = -C \int dk_y  \frac{k_y r_0^2 q^\prime}{q^2}  \tau_{c,k} S(k_y)\cdot
	\int\limits^{x=w_k}_0 dx\frac{  F(x/w_k)}{1 + (x/x_s)^2} 
	\bigg( \rho  c_s^2  \frac{\partial}{\partial x} \langle u_z \rangle  \bigg)
\end{split}
\end{equation}
We ignore the $(x/x_s)^2$ in the denominator for that in the integration of step function $F(x/w_k)$, the $1/(1+x^2/x_s^2 )\rightarrow 1$ (see \Fref{fig:Material_8}).
Hence, in this limit, we obtain
\begin{equation}
	\int\limits^{x=w_k}_0 dx\frac{  F(x/w_k)}{1 + (x/x_s)^2}  
	\simeq \int\limits^{x=w_k}_0 dx F(x/w_k)
	=1
\end{equation}
Hence, we have
\begin{equation}
\begin{split}
	\textcircled{a}
	&= C
	\int dk_y  \frac{k_y r_0^2 q^\prime}{q^2}  \tau_{c,k}  S(k_y) 
	 \cdot
	\bigg( -\rho  c_s^2  \frac{\partial}{\partial x} \langle u_z \rangle  \bigg)
	\\
	&=\frac{b_{x,0}^2 \cancel{\int dk_y   \frac{k_y r_0^2 q^\prime}{q^2}S(k_y) } \tau_{c,k} }{
	\cancel{\int dk_y  \frac{k_y r_0^2 q^\prime}{q^2}  S(k_y) } }   \bigg( -\rho  c_s^2  \frac{\partial}{\partial x} \langle u_z \rangle  \bigg)
	\\
	& =-\rho  c_s^2 \sum \limits_{k_y k_z} |\widetilde{b}_{x,k}|^2 \tau_{c,k}  \frac{\partial}{\partial x} \langle u_z \rangle	
\end{split}
\end{equation}
Also, the second term in \Eref{eq: bp_2} becomes
\begin{equation}
\begin{split}
	\textcircled{b} 
	&= 
	\sum\limits_{k_y k_z} |\widetilde{b}_{x,k}|^2 \frac{1}{(k_\perp^2 D_T)^2 + k_z^2 c_s^2}  
	\bigg( i k_z c_s^2  \frac{\partial}{\partial x} \langle p \rangle 
	\bigg)
	\\
	& = -i \sum\limits_{k_y k_z} \frac{|\widetilde{b}_{x,k}|^2}{k_z} (-1 + \frac{k_\perp^4 D_T^4}{k_\perp^4 D_T^4 + k_z^2 c_s^2}) \bigg( \frac{\partial }{\partial x} \langle p\rangle  \bigg),
\end{split}
\end{equation}
In this limit, we have $k_\perp^2 D_T \gg k_z^2 c_s^2$ such that 
\begin{equation}
	-1 + \frac{k_\perp^4 D_T^4}{k_\perp^4 D_T^4 + \cancel{k_z^2 c_s^2}} \simeq 0.
\end{equation}
Hence, the second term can be approximated as $\textcircled{b}  \simeq 0$, and $\langle \widetilde{b}_x \widetilde{p} \rangle$ can be simplified to 
\begin{equation}
	\langle \widetilde{b}_x \widetilde{p} \rangle 
	\simeq -\rho  c_s^2 \sum  \limits_{k_yk_z} |\widetilde{b}_{x,k}|^2 \tau_{c,k}  \frac{\partial}{\partial x} \langle u_z \rangle.
\end{equation}

\section{Weak Turbulence Limit} \label{appen:Weak trub}
In weak fluid turbulence, we have $ w_k \gg x_s$ ( or $k_\parallel c_s > k_\perp^2 D_T  $). 
The integral in equation (\ref{eq: circled a_2 })  becomes
\begin{equation}
\begin{split}
	\textcircled{a}
	&=\sum\limits_{k_y k_z} |\widetilde{b}_{x,k}|^2 \frac{\tau_{c,k}}{1 + (k_y x/L_s)^2 c_s^2 \tau_{c,k}^2} 
	\bigg( -\rho  c_s^2  \frac{\partial}{\partial x} \langle u_z \rangle  \bigg)
	\\
	 &=  
	  -C \int dk_y  \frac{k_y r_0^2 q^\prime}{q^2}  S(k_y) \int dx F(x/w_k)\frac{\tau_{c,k}}{1 + (x/x_s)^2}
	  \cdot 
	\bigg( \rho  c_s^2  \frac{\partial}{\partial x} \langle u_z \rangle  \bigg)
	\\
	&\simeq 
	-\frac{b_{x,0}^2 \cancel{\int dk_y  \frac{k_y r_0^2 q^\prime}{q^2}  S(k_y)} }{\cancel{\int dk_y \frac{k_y r_0^2 q^\prime}{q^2}  S(k_y) }}
	\cdot 
	\tau_{c,k} \frac{x_s}{w_k} 
	F(0)
	 \cdot
	 \underbrace{\int\limits^{x =x_s}_0 
	 \frac{  d(x/x_s)}{1 + (x/x_s)^2} }_{=arctan(x_s/x_s) = \pi/4}
	 \cdot
	 \bigg( \rho  c_s^2  \frac{\partial}{\partial x} \langle u_z \rangle  \bigg)
\end{split}
\end{equation}
where $\arctan{(x_s/x_s)} = \arctan{(1)} =\pi/4$ and $\tau_{d,k} \simeq   L_s /k_y c_s w_k$ is dispersal timescale of acoustic packet propagating along stochastic magnetic fields. 
Hence,  
\begin{equation}
	\textcircled{a}
	\simeq
	-b_{x,0}^2 \tau_{d,k} (\frac{x_s}{w_k})F(0)\frac{\pi}{4}\bigg( \rho  c_s^2  \frac{\partial}{\partial x} \langle u_z \rangle  \bigg).
\end{equation}
Note that this term scales $\propto x_s/w_k$, which is asymptotically small as $x_s/w_k \rightarrow 0$.
Then $\textcircled{a} \rightarrow 0$, so the first term is negligible. 
The second term in equation (\ref{eq: circled a_2 }) becomes
\begin{equation}
\begin{split}
	\textcircled{b} 
	&= \sum\limits_{k_y k_z} |\widetilde{b}_{x,k}|^2 \frac{1}{(k_\perp^2 D_T)^2 + k_z^2 c_s^2}  
	\big( i k_z c_s^2  \frac{\partial}{\partial x} \langle p \rangle 
	\big)
\\
	&=  \sum\limits_{k_y k_z} |\widetilde{b}_{x,k}|^2 \frac{-i}{k_z}
	(-1 + \underbrace{\frac{(k_\perp^2 D_T)^2}{(k_\perp^2 D_T)^2 + k_z^2 c_s^2} }_{=0} )
	\big(   \frac{\partial}{\partial x} \langle p \rangle 
	\big),
\end{split}
\end{equation}
where the response term is approximated as $i/k_z$, since in this limit turbulent scattering is weak---i.e. $k_\perp^2 D_T \rightarrow 0$.
So, we obtain
\begin{equation}
\begin{split}
	\textcircled{b} 
	&= \sum  \limits_{k_yk_z} |\widetilde{b}_{x,k}|^2 \frac{1}{(k_\perp^2 D_T)^2 + k_z^2 c_s^2}  
	\big( i k_z c_s^2  \frac{\partial}{\partial x} \langle p \rangle 
	\big)
	\\
	&\simeq  \sum  \limits_{k_yk_z} |\widetilde{b}_{x,k}|^2 \frac{i}{k_z}
	   \frac{\partial}{\partial x} \langle p \rangle  .
	   \end{split}
\end{equation} 
Here, $\sum\limits_{k_z} i/k_z $ can be approximate as 
\begin{equation}
\begin{split}
	\sum\limits_{k_z} \frac{i}{k_z -i\delta} 
	&=
	\sum\limits_{k_z} i PV\bigg[ \frac{1}{k_z} \bigg] - \pi \delta(k_z)
	\\
	& = 0 - \pi \delta(k_z ) ,
\end{split}
\end{equation}
where $PV$ is Cauchy principle value and $\pi \delta(k_z c_s) $ as $\tau_{d,k}$.
So, we have
\begin{equation}
\begin{split}
	\textcircled{b}
	  & \simeq  - \sum  \limits_{k_yk_z} |\widetilde{b}_{x,k}|^2
	 \pi \delta(k_z)
	   \frac{\partial}{\partial x} \langle p \rangle
	   \\
	   & \simeq-D_M \frac{\partial}{\partial x} \langle p \rangle,
\end{split}
\end{equation}  
where $D_M$ is the magnetic diffusivity.
Hence, the kinetic stress flux is 
\begin{equation}
	\langle \widetilde{b}_x \widetilde{p} \rangle 
	=
	-b_{x,0}^2 \tau_{d,k} F(0)\frac{\pi}{4} \rho  c_s^2  \frac{\partial}{\partial x} \langle u_z \rangle 
	 -D_M	  \frac{\partial}{\partial x} \langle p \rangle .
	\label{eq: p perturb. in weak turb.}
\end{equation}
The first term on RHS is approximate $\textcircled{a} \simeq 0$ for $F(0) \simeq 0$ in this limit.
So, we obtain
. \begin{equation}
	\langle \widetilde{b}_x \widetilde{p} \rangle 
	= 
	 -D_M	  \frac{\partial}{\partial x} \langle p \rangle 
\end{equation}
in this limit.

\section*{References}
\bibliographystyle{unsrt}

\bibliography{main.bib}

\end{document}